\documentclass[sigconf,nonacm]{acmart}

\usepackage{subfigure}
\usepackage{siunitx}
\usepackage{enumitem}

\usepackage{algorithm}
\usepackage[noend]{algpseudocode}
\let\ReturnInline\Return
\renewcommand{\Return}{\State\ReturnInline}
\algrenewcommand\algorithmicrequire{$\rhd$}
\algrenewcommand\algorithmicensure{$\square$}

\AtBeginDocument{%
  \providecommand\BibTeX{{%
    \normalfont B\kern-0.5em{\scshape i\kern-0.25em b}\kern-0.8em\TeX}}}

\setcopyright{acmcopyright}
\copyrightyear{2018}
\acmYear{2018}
\acmDOI{XXXXXXX.XXXXXXX}

\acmConference[Conference acronym 'XX]{Make sure to enter the correct
  conference title from your rights confirmation emai}{June 03--05,
  2018}{Woodstock, NY}
\newcommand{\ignore}[1]{}

\begin{document}

%% Full title of the paper.
\title[CPU vs. GPU for Community Detection: Performance Insights from GVE-Louvain and $\nu$-Louvain]{CPU vs. GPU for Community Detection: \\Performance Insights from GVE-Louvain and $\nu$-Louvain}

%% Short title to be used in page headers (optional).
% \title[short title]{full title}
% \subtitle{Something other than the title}

%% Authors and their affiliations.
\author{Subhajit Sahu}
\email{subhajit.sahu@research.iiit.ac.in}
\affiliation{%
  \institution{IIIT Hyderabad}
  \streetaddress{Professor CR Rao Rd, Gachibowli}
  \city{Hyderabad}
  \state{Telangana}
  \country{India}
  \postcode{500032}
}

%% Concise author list in page headers.
%\renewcommand{\shortauthors}{Sahu, Kothapalli, and Banerjee, et al.}

%% Show page numbers.
\settopmatter{printfolios=true}

%% Short summary of the work to be presented in the article.
\begin{abstract}
Community detection involves identifying natural divisions in networks, a crucial task for many large-scale applications. This report presents GVE-Louvain, one of the most efficient multicore implementations of the Louvain algorithm, a high-quality method for community detection. Running on a dual 16-core Intel Xeon Gold 6226R server, GVE-Louvain outperforms Vite, Grappolo, NetworKit Louvain, and cuGraph Louvain (on an NVIDIA A100 GPU) by factors of $50\times$, $22\times$, $20\times$, and $5.8\times$, respectively, achieving a processing rate of $560M$ edges per second on a $3.8B$-edge graph. Additionally, it scales efficiently, improving performance by $1.6\times$ for every thread doubling. The paper also presents $\nu$-Louvain, a GPU-based implementation. When evaluated on an NVIDIA A100 GPU, $\nu$-Louvain performs only on par with GVE-Louvain, largely due to reduced workload and parallelism in later algorithmic passes. These results suggest that CPUs, with their flexibility in handling irregular workloads, may be better suited for community detection tasks.
\end{abstract}

%% The code below is generated by the tool at http://dl.acm.org/ccs.cfm.

% \ccsdesc[500]{Theory of computation~Parallel algorithms}
% \ccsdesc[500]{Theory of computation~Graph algorithms analysis}

%% Pick words that accurately describe the work being presented.
% \keywords{Community detection \and Louvain algorithm \and Multicore algorithm \and Suitability of CPUs.}

% \received{20 February 2007}
% \received[revised]{12 March 2009}
% \received[accepted]{5 June 2009}

%% Process the author and title information.
\maketitle

\section{Introduction}
\label{sec:introduction}
Community detection is a fundamental task in network analysis, involving the identification of groups of vertices (communities) that are more densely connected internally than with the rest of the network \cite{com-fortunato10}. This problem is tackled in an unsupervised manner, without prior knowledge of the number or size distribution of communities \cite{com-blondel08}. The identified communities provide valuable insights into the structure and behavior of complex systems \cite{com-fortunato10, abbe2018community}, with applications spanning fields such as ecology \cite{guimera2010origin}, machine learning \cite{bai2024leiden, das2011unsupervised}, healthcare \cite{salathe2010dynamics, bechtel2005lung, haq2016community}, and neuroscience \cite{he2010graph}, among others. When communities are derived solely from network topology, they are termed intrinsic, and if vertices belong to only one community, the divisions are considered disjoint \cite{com-gregory10, coscia2011classification}. The modularity metric is commonly used to evaluate the identified communities \cite{com-newman06}.

The widespread utility of community detection has driven the development of numerous algorithms \cite{com-raghavan07, com-blondel08, com-xie11, traag2023large, clauset2004finding, duch2005community, reichardt2006statistical, com-kloster14, com-traag19, com-you20, com-rosvall08, com-whang13}, with the Louvain algorithm \cite{com-blondel08}, a hierarchical and greedy approach to modularity optimization, emerging as a popular choice due to its ability to identify high-quality communities \cite{com-newman06}. Consequently, several studies on Louvain have been conducted, which propose\ignore{a number of algorithmic} several optimizations \cite{com-rotta11, com-waltman13, com-gach14, com-traag15, com-lu15, com-ryu16, com-ozaki16, com-naim17, com-halappanavar17, com-ghosh18, com-traag19, com-zhang21, com-shi21, com-you22, com-aldabobi22} and parallelization techniques \cite{com-cheong13, com-wickramaarachchi14, com-lu15, com-zeng15, com-que15, com-naim17, com-fazlali17, com-halappanavar17, com-zeitz17, com-ghosh18, com-bhowmik19, com-gheibi20, com-shi21, com-bhowmick22}. Further, significant research effort has been focused on developing efficient parallel implementations of Louvain algorithm for multicore CPUs \cite{staudt2015engineering, staudt2016networkit, com-fazlali17, com-halappanavar17, qie2022isolate}, GPUs \cite{com-naim17}, CPU-GPU hybrids \cite{com-bhowmik19, com-mohammadi20}, multi-GPUs \cite{com-cheong13, kang2023cugraph, chou2022batched, com-gawande22}, and multi-node systems \cite{com-ghosh18, ghosh2018scalable, sattar2022scalable, com-bhowmick22}\ignore{--- CPU only \cite{com-ghosh18, ghosh2018scalable, sattar2022scalable} and CPU-GPU hybrids \cite{com-bhowmick22}}.

However, many of the aforementioned works focus on optimizing the local-moving phase of the Louvain algorithm but neglect the aggregation phase, which becomes a bottleneck once the local-moving phase is optimized. In addition, while much research has explored GPU-based solutions, developing efficient GPU algorithms poses challenges in both implementation and maintenance, compounded by the rising costs of GPUs. In contrast, the multicore/shared memory environment is well-suited for community detection due to its energy efficiency and the widespread availability of hardware with ample DRAM. In this technical report, we present one of the most efficient multicore implementations of the Louvain algorithm, GVE-Louvain\footnote{\url{https://github.com/puzzlef/louvain-communities-openmp}}. Furthermore, we also present our GPU implementation of the Louvain method, $\nu$-Louvain\footnote{\url{https://github.com/puzzlef/louvain-communities-cuda}}. However, running on an NVIDIA A100 GPU, it does not necessarily outperform GVE-Louvain, suggesting that CPUs, with their ability to handle irregular workloads, may be better suited for community detection tasks. A comparison of GVE-Louvain against other state-of-the-art implementations is given in Table \ref{tab:compare}.

\begin{table}[hbtp]
  \centering
  \caption{Speedup of GVE-Louvain, our multicore implementation of the Louvain algorithm, compared to other state-of-the-art multicore and GPU-based implementations.}
  \label{tab:compare}
  \begin{tabular}{|c|c||c|}
    \toprule
    \textbf{Louvain implementation} &
    \textbf{Parallelism} &
    \textbf{Our Speedup} \\
    \midrule
    Vite (1 node) \cite{ghosh2018scalable} & Multi node & $50\times$ \\ \hline
    Grappolo \cite{com-halappanavar17} & Multicore & $22\times$ \\ \hline
    NetworKit Louvain \cite{staudt2016networkit} & Multicore & $20\times$ \\ \hline
    Nido (1 GPU) \cite{chou2022batched} & Multi GPU & $56\times$ \\ \hline
    cuGraph Louvain (1 GPU) \cite{kang2023cugraph} & Multi GPU & $5.8\times$ \\ \hline
  \bottomrule
  \end{tabular}
\end{table}

\section{Related work}
\label{sec:related}
The \textit{Louvain} method, introduced by Blondel et al. \cite{com-blondel08} from the University of Louvain, is a popular community detection algorithm based on greedy modularity optimization. Known for identifying communities with high modularity, it has garnered significant attention \cite{com-lancichinetti09}. Several algorithmic enhancements have been proposed to improve its efficiency, including early pruning of non-promising candidates such as leaf vertices \cite{com-ryu16, com-halappanavar17, com-zhang21, com-you22}, restricting local moves to likely vertices \cite{com-ryu16, com-ozaki16, com-zhang21, com-shi21}, and ordering vertices based on node importance \cite{com-aldabobi22}. Other modifications include moving nodes to a random neighbor community \cite{com-traag15}, threshold scaling \cite{com-lu15, com-naim17, com-halappanavar17}, threshold cycling \cite{com-ghosh18}, subnetwork and multilevel refinements \cite{com-waltman13, com-traag19, com-rotta11, com-gach14, com-shi21}, and early termination \cite{com-ghosh18}.

Several strategies have been proposed to parallelize the Louvain algorithm. These include employing heuristics to overcome the sequential nature of the algorithm \cite{com-lu15}, ordering vertices using graph coloring \cite{com-halappanavar17}, performing iterations asynchronously \cite{com-que15, com-shi21}, using adaptive parallel thread assignment \cite{com-fazlali17, com-naim17, com-sattar19, com-mohammadi20}, parallelizing the computationally intensive first iteration \cite{com-wickramaarachchi14}, leveraging vector-based hash tables \cite{com-halappanavar17}, and replacing hashing with sort-reduce techniques \cite{com-cheong13}. It is important to note, however, that community detection methods relying on modularity maximization, such as Louvain, are known to suffer from the resolution limit problem, which restricts identification of communities of certain sizes \cite{com-ghosh19}.

We now discuss state-of-the-art implementations of Parallel Louvain. Ghosh et al. \cite{ghosh2018scalable} propose Vite, a distributed memory parallel implementation of the Louvain method, incorporating heuristics to improve performance. Similarly, Grappolo, by Halappanavar et al. \cite{com-halappanavar17}, is a shared memory parallel implementation. NetworKit \cite{staudt2016networkit} is a hybrid software package for analyzing large graph datasets, combining C++ kernels with a Python frontend and includes a parallel Louvain implementation. Chou and Ghosh \cite{chou2022batched} present Nido, a batched clustering method for GPUs that processes graphs larger than a node's combined GPU memory. Finally, cuGraph \cite{kang2023cugraph} is a GPU-accelerated graph analytics library within the RAPIDS suite, leveraging NVIDIA GPUs to accelerate graph analytics. Written in C++ with CUDA, cuGraph is primarily accessed through Python, making it user-friendly for data scientists and developers.

\section{Preliminaries}
\label{sec:preliminaries}
Let $G(V, E, w)$ be an undirected graph where $V$ is the set of vertices, $E$ is the set of edges, and $w_{ij} = w_{ji}$ represents the weight of each edge (if $G$ is unweighted, then $w_{ij} = 1$). The neighbors of a vertex $i$ are given by $J_i = \{j \mid (i, j) \in E\}$, and the weighted degree of $i$ is $K_i = \sum_{j \in J_i} w_{ij}$. The graph contains $N = |V|$ vertices, $M = |E|$ edges, and the total edge weight is $m = \sum_{i, j \in V} w_{ij} / 2$.

\subsection{Community detection}
\label{sec:about-communities}

Disjoint community detection assigns each vertex $i \in V$ to a community $c$ from a set $\Gamma$ using a community membership function $C: V \to \Gamma$. The vertices in community $c$ are denoted by $V_c$, and the community of vertex $i$ is $C_i$. For a vertex $i$, its neighbors in community $c$ are given by $J_{i \to c} = \{j \mid j \in J_i \text{ and } C_j = c\}$, and the sum of their edge weights is $K_{i \to c} = \sum_{j \in J_{i \to c}} w_{ij}$. The total weight of edges within community $c$ is $\sigma_c = \sum_{(i, j) \in E \text{ and } C_i = C_j = c} w_{ij}$, and the total edge weight associated with $c$ is $\Sigma_c = \sum_{(i, j) \in E \text{ and } C_i = c} w_{ij}$.

\subsection{Modularity}
\label{sec:about-modularity}

Modularity is a fitness metric used to evaluate the quality of communities detected by heuristic-based algorithms. It is calculated as the difference between the fraction of edges within communities and the expected fraction if edges were randomly distributed. It ranges from $[-0.5, 1]$, with higher values indicating better outcomes \cite{com-brandes07}. The modularity $Q$ of detected communities is computed using Equation \ref{eq:modularity}. Furthermore, the \textit{delta modularity} $\Delta Q_{i: d \rightarrow c}$, which quantifies the change in modularity when transferring a vertex $i$ from community $d$ to $c$, is calculated using Equation \ref{eq:delta-modularity}.

\begin{equation}
\label{eq:modularity}
  Q
  = \sum_{c \in \Gamma} \left[\frac{\sigma_c}{2m} - \left(\frac{\Sigma_c}{2m}\right)^2\right]
\end{equation}

\begin{equation}
\label{eq:delta-modularity}
  \Delta Q_{i: d \rightarrow c}
  = \frac{1}{m} (K_{i \rightarrow c} - K_{i \rightarrow d}) - \frac{K_i}{2m^2} (K_i + \Sigma_c - \Sigma_d)
\end{equation}

\subsection{Louvain algorithm}
\label{sec:about-louvain}

The Louvain method \cite{com-blondel08} is an agglomerative algorithm for modularity optimization that identifies high-quality disjoint communities in large networks. It has a time complexity of $O(L|E|)$, where $L$ is the total number of iterations, and a space complexity of $O(|V| + |E|)$ \cite{com-lancichinetti09}. The algorithm operates in two phases: the \textit{local-moving phase} and the \textit{aggregation phase}. During the local-moving phase, each vertex $i$ greedily moves to the community of a neighboring vertex $j \in J_i$ that maximizes the increase in modularity $\Delta Q_{i:C_i \rightarrow C_j}$, calculated using Equation \ref{eq:delta-modularity}. In the aggregation phase, all vertices in a community are merged into a single super-vertex. These phases constitute one pass of the algorithm, which repeats until no further modularity increase is achieved \cite{com-blondel08}.

\subsection{Open Addressing in Hashing}

Open addressing, or closed hashing, is a collision resolution technique in hash tables that stores all entries within a single array, avoiding auxiliary structures like linked lists \cite{tenenbaum1990data}. Collisions are resolved through probing, where alternative positions --- determined by a probe sequence --- are searched until the key is found or an empty slot appears. Common probe sequences include: \textbf{(1)} \textit{Linear probing}, which uses a fixed step size (often $1$); \textbf{(2)} \textit{Quadratic probing}, where probe intervals grow quadratically; and \textbf{(3)} \textit{Double hashing}, which determines step sizes using a second hash function. Linear probing has optimal cache performance but suffers from clustering, double hashing minimizes clustering at the cost of weaker cache performance, and quadratic probing balances both. The load factor, the ratio of occupied slots to total capacity, critically impacts efficiency, with performance degrading as it nears $100\%$. A typical load factor for open addressing is around $50\%$.

\subsection{Fundamentals of a GPU}

Streaming Multiprocessors (SMs) are the fundamental building block of NVIDIA GPUs. Each SM contains multiple CUDA cores, shared memory, registers, and specialized function units. The number of SMs varies by GPU model, with each operating independently. The memory hierarchy of a GPU includes global memory (large but slow), shared memory (low-latency, shared within an SM), and local memory (private to threads when registers are insufficient). Threads are organized into warps (32 threads executing in lockstep), thread blocks (groups of threads running on the same SM), and grids (collections of thread blocks). SMs schedule warps dynamically to optimize performance, switching execution when threads stall. Threads within a block use shared memory for fast communication, while global memory enables data exchange across thread blocks, albeit with higher latency \cite{cuda-sanders10, gpu-nickolls10}.

\section{Approach}
\label{sec:approach}
\subsection{Optimizations for Multicore Louvain\ignore{algorithm}}
\label{sec:louvain}

We implement a parallel version of the Louvain method to experiment with various techniques and fine-tune parameter settings. Our implementation uses an asynchronous approach where threads independently process different parts of the graph. This enhances convergence speed but introduces variability in the final results \cite{com-blondel08, com-halappanavar17}. Additionally, we allocate a separate hashtable for each thread to track the delta-modularity when moving to each community linked to a vertex during the local-moving phase, and to record the total edge weight between super-vertices in the aggregation phase.

\begin{figure*}[hbtp]
  \centering
  \subfigure{
    \label{fig:louvain-opt--all}
    \includegraphics[width=0.98\linewidth]{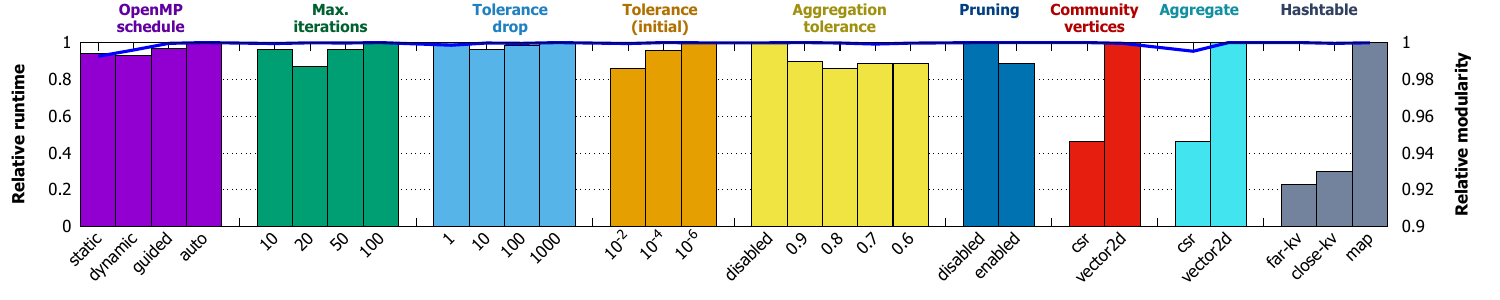}
  } \\[-2ex]
  \caption{Impact of various parameter controls and optimizations on the runtime and result quality (modularity) of the Louvain algorithm, with the left/right Y-axes showing the effect of each optimization on relative runtime/modularity, respectively.}
  \label{fig:louvain-opt}
\end{figure*}

For each optimization, we attempt a number of alternatives and compare their relative time and modularity of the resulting communities, as shown in Figure \ref{fig:louvain-opt}. These results are based on tests run five times per graph in the dataset (Table \ref{tab:dataset}) to minimize noise. The geometric mean is used for runtime, while the arithmetic mean is used for modularity --- expressed as ratios within each category.

\ignore{Our optimizations include using OpenMP's \verb|dynamic| loop schedule, limiting the number of iterations per pass to $20$, using a tolerance drop rate of $10$, setting an initial tolerance of $0.01$, using an aggregation tolerance of $0.8$, employing vertex pruning, making use of parallel prefix sum and preallocated Compressed Sparse Row (CSR) data structures for finding community vertices and for storing the super-vertex graph during the aggregation phase, and using fast collision-free per-thread hashtables which are well separated in their memory addresses (\textit{Far-KV}) for the local-moving and aggregation phases of the algorithm. Details\ignore{ on each of the optimizations is} are given below.}

\subsubsection{Adjusting OpenMP loop schedule}

We evaluate OpenMP's \textit{static}, \textit{dynamic}, \textit{guided}, and \textit{auto} loop scheduling approaches to parallelize the local-moving and aggregation phases of the Louvain algorithm --- each with a chunk size of $2048$ (using smaller chunk sizes would improve load balance, but can significantly increase the scheduling cost \cite{sahu2024lockfree}). Our results show that while scheduling has a small impact on the quality of detected communities, OpenMP's \texttt{dynamic} loop schedule is the best choice. It improves load balancing in cases of non-uniform vertex degree distribution and reduces runtime by $7\%$ compared to the \textit{auto} schedule, with only a negligible $0.4\%$ decrease in modularity, likely attributable to noise.

\subsubsection{Limiting the number of iterations per pass}

Restricting the number of iterations in the local-moving phase ensures timely termination, which is crucial since this phase in the first pass is the most computationally expensive step of the algorithm. However, setting the limit too low can negatively impact the convergence rate. Our results show that capping iterations at $20$ improves convergence speed by $13\%$ compared to a limit of $100$.

\subsubsection{Adjusting tolerance drop rate (threshold scaling)}

Tolerance in the Louvain algorithm determines convergence during the local-moving phase, which occurs when the total delta-modularity $\Delta Q$ in an iteration falls below or equals the specified tolerance $\tau$. Instead of maintaining a fixed tolerance across all passes, we can apply threshold scaling --- starting with a high initial tolerance and gradually reducing it \cite{com-lu15, com-naim17, com-halappanavar17}. Our findings show that a tolerance drop rate of $10$ improves convergence speed by $4\%$ compared to a drop rate of $1$ (i.e., with threshold scaling disabled), without compromising the quality of detected communities.

\subsubsection{Adjusting initial tolerance}

Using a smaller initial tolerance enables the algorithm to explore a wider range of community assignments early on but increases runtime. We find that setting the initial tolerance $\tau$ to $0.01$ reduces runtime by $14\%$ without compromising the quality of communities, compared to using $\tau = 10^{-6}$.

\subsubsection{Adjusting aggregation tolerance}
\label{sec:adjust-aggregation-tolerance}

In later passes of the algorithm, we observe that too few communities merge, yet the aggregation phase is expensive. To address this, we introduce the aggregation tolerance $\tau_{agg}$, which determines when communities are considered to have converged based on the number of merges. If too few communities merge in a given pass (i.e., $|V_{aggregated}|/|V| \geq \tau_{agg}$), the algorithm stops, enabling earlier termination when additional merges have minimal impact on the final result. Our experiments show that an aggregation tolerance of $\tau_{agg} = 0.8$ is optimal, reducing runtime by $14\%$ compared to disabling it ($\tau_{agg} = 1$) while maintaining equivalent community quality.

\begin{figure}[hbtp]
  \centering
  \subfigure{
    \label{fig:louvain-pruning--all}
    \includegraphics[width=0.58\linewidth]{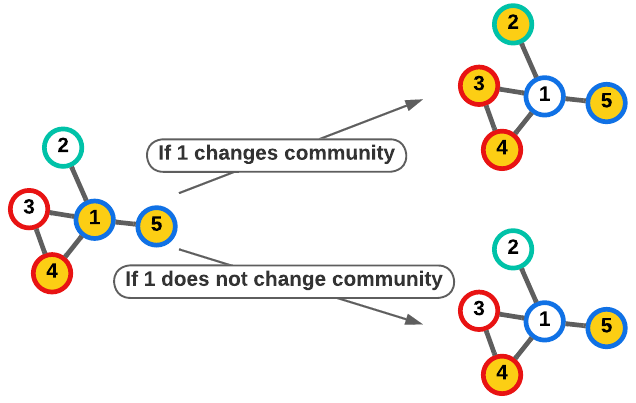}
  } \\[-2ex]
  \caption{Illustration of vertex pruning optimization: After processing vertex $1$, it's unmarked. If vertex $1$ changes its community, its neighbors are marked for processing. Community membership of each vertex is depicted by border color, and marked vertices are highlighted in yellow.}
  \label{fig:louvain-pruning}
\end{figure}

\subsubsection{Vertex pruning}

Vertex pruning \cite{com-ryu16, com-ozaki16, com-zhang21, com-shi21, com-ghosh18, zhang2018self, tithi2020prune} is a technique used to minimize unnecessary computation by marking neighbors for processing when a vertex changes its community. Once processed, a vertex is marked as no longer needing further processing. While this approach introduces the overhead of marking and unmarking vertices, our results show that the performance improvement justifies the cost, with a performance boost of $11\%$. An illustration of this optimization is shown in Figure \ref{fig:louvain-pruning}.

\subsubsection{Finding community vertices for aggregation phase}

In the aggregation phase of the Louvain algorithm, communities from the previous local-moving phase are combined into super-vertices, with edges between super-vertices reflecting the total weight of edges between the respective communities. To achieve this, we need the list of vertices in each community, rather than the community membership of individual vertices. A straightforward approach would use two-dimensional arrays to store vertices for each community, where the first dimension represents the community ID and the second points to the $n^{th}$ vertex in that community. However, this approach requires costly memory allocation during the algorithm. Using a parallel prefix sum technique with a preallocated Compressed Sparse Row (CSR) data structure avoids repeated memory allocation, improving performance. Our findings show that this method is $2.2\times$ faster than using 2D arrays for aggregating vertices.

\subsubsection{Storing aggregated communities (super-vertex graph)}

Once the list of vertices in each community is obtained, the communities are aggregated into super-vertices, with edges between super-vertices representing the total weight of edges between the respective communities. This resulting structure is known as the super-vertex graph, or compressed graph, and is used as input for the local-moving phase of the next pass in the Louvain algorithm. To store the super-vertex graph efficiently, a two-dimensional array in the adjacency list format can be used, but this requires memory allocation during the algorithm, which negatively impacts performance. A more efficient approach involves using two preallocated CSRs --- one for the source graph and one for the target graph (except in the first pass, where the input graph may be stored in any desired format, such as one that is suitable for dynamic batch updates) --- along with parallel prefix sum. This improves performance, making the process $2.2\times$ faster compared to using 2D arrays.

\subsubsection{Hashtable design for local-moving/aggregation phases}

\begin{figure}[hbtp]
  \centering
  \subfigure{
    \label{fig:louvain-hashtable--all}
    \includegraphics[width=0.78\linewidth]{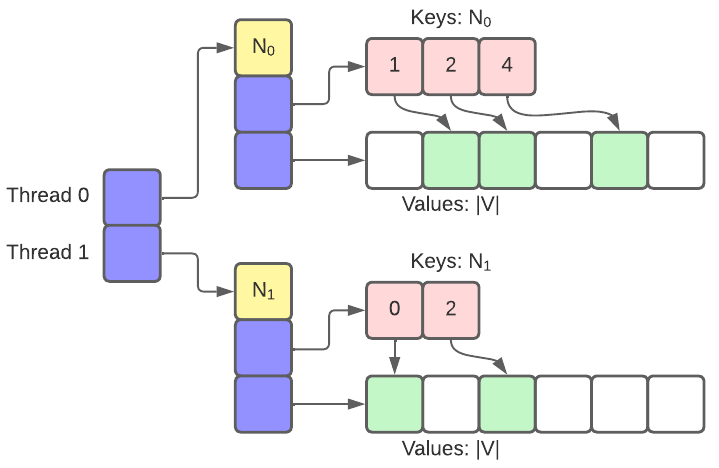}
  } \\[-2ex]
  \caption{Illustration of collision-free per-thread hashtables that are well separated in their memory addresses (Far-KV), for two threads. Each hashtable consists of a keys vector, a values vector of size $|V|$, and a key count ($N_0$ or $N_1$). The value associated with each key is stored or accumulated at the index pointed to by the key. To prevent false cache sharing, the key count for each hashtable is independently updated and allocated separately on the heap. These hashtables are utilized during the local-moving and aggregation phases of our multicore Louvain implementation, GVE-Louvain.}
  \label{fig:louvain-hashtable}
\end{figure}

C++'s inbuilt maps can be used as per-thread independent hashtables for the Louvain algorithm, but they offer poor performance. To improve this, we use a key-list and a collision-free full-size values array, significantly boosting performance. However, when the memory addresses of the hashtables are close in memory (\textit{Close-KV}), as seen in NetworKit Louvain \cite{staudt2016networkit}, performance still suffers, even with exclusive thread usage. This may be due to false cache-sharing. On the other hand, when the hashtable memory addresses are farther apart (\textit{Far-KV}), performance improves. Our results show that \textit{Far-KV} outperforms both \textit{Map} by $4.4\times$ and \textit{Close-KV} by $1.3\times$. An illustration of the \textit{Far-KV} hashtable is shown in Figure \ref{fig:louvain-hashtable}.

\subsection{Our optimized Multicore Louvain implementation}

\begin{figure*}[hbtp]
  \centering
  \subfigure{
    \label{fig:louvain-pass--all}
    \includegraphics[width=0.98\linewidth]{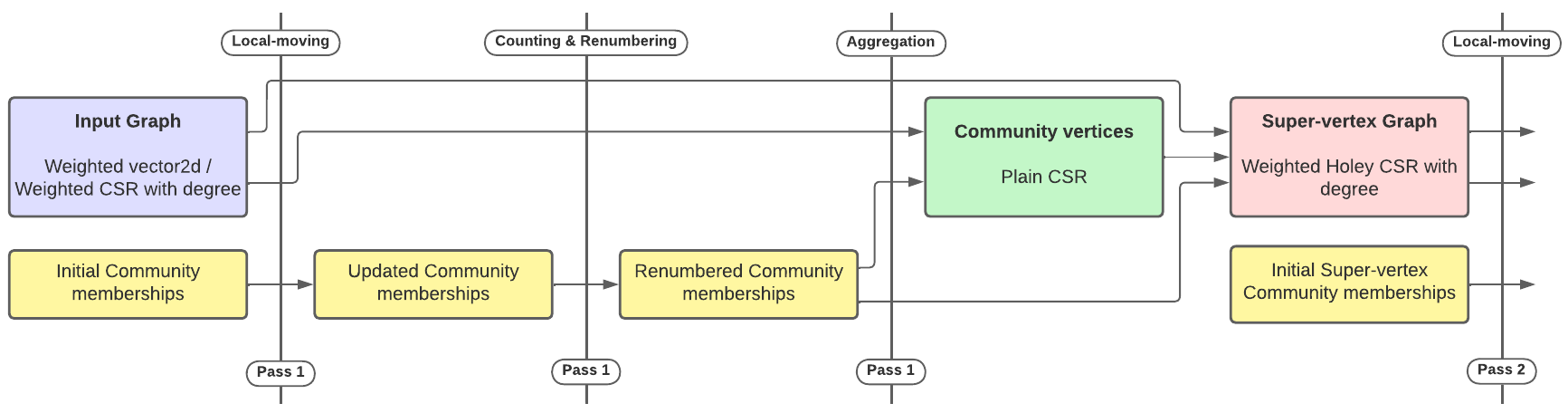}
  } \\[-2ex]
  \caption{A flow diagram illustrating the first pass of GVE-Louvain for a Weighted 2D-vector based or a Weighted CSR with degree based input graph. In the local-moving phase, vertex community memberships are updated until the total change in delta-modularity across all vertices reaches a specified threshold. Community memberships are then counted and renumbered. In the aggregation phase, community vertices in a CSR are first obtained. This is used to create the super-vertex graph stored in a Weighted Holey CSR with degree. In subsequent passes, the input is a Weighted Holey CSR with degree and initial community membership for super-vertices from the previous pass.}
  \label{fig:louvain-pass}
\end{figure*}

We now explain the implementation of GVE-Louvain, whose psuedocode is given in Algorithms \ref{alg:louvain}, \ref{alg:louvainlm}, and \ref{alg:louvainag}. We aim to incorporate GVE-Louvain into our upcoming command-line graph processing tool named "GVE", short for "Graph(Vertices, Edges)". GVE-Louvain has a time complexity of $O(KM)$, where $K$ is the total number of iterations, and a space complexity of $O(TN + M)$, where $T$ is the number of threads used, and $TN$ represents space utilized by the collision-free hash tables $H_t$ for each thread. A flow diagram illustrating the first pass of GVE-Louvain is shown in Figure \ref{fig:louvain-pass}.

\subsubsection{Main step of GVE-Louvain}

The main step of GVE-Louvain, implemented in the \texttt{louvain()} function, is outlined in Algorithm \ref{alg:louvain}. It involves initialization, the local-moving phase, and the aggregation phase. The \texttt{louvain()} function takes the input graph $G$ and returns the community membership $C$ for each vertex. In line \ref{alg:louvain--initialization}, we initialize the community membership $C$ for each vertex in $G$ and perform multiple passes of the Louvain algorithm, limited to $MAX\_PASSES$ (lines \ref{alg:louvain--passes-begin}-\ref{alg:louvain--passes-end}). During each pass, we reset the total edge weight of each vertex $K'$, the total edge weight of each community $\Sigma'$, and the community membership $C'$ of each vertex in the current graph $G'$ (line \ref{alg:louvain--reset-weights}), and mark all vertices as unprocessed (line \ref{alg:louvain--reset-affected}). Next, the local-moving phase is executed by calling \texttt{louvainMove()} (Algorithm \ref{alg:louvainlm}), which optimizes community assignments (line \ref{alg:louvain--local-move}). If the local-moving phase converges in a single iteration (indicated by a modularity change $\Delta Q \leq \tau$, where $\tau$ is the iteration tolerance), the algorithm stops, having reached global convergence (line \ref{alg:louvain--globally-converged}). The algorithm can also terminate early if the drop in community count is too small (line \ref{alg:louvain--aggregation-tolerance}), i.e., $|\Gamma|/|\Gamma_{old}| > \tau_{agg}$, where $|\Gamma_{old}|$ is the number of communities before the local-moving phase (in the current pass) and $|\Gamma|$ is the number after, with $\tau_{agg}$ being the aggregation tolerance (see Section \ref{sec:adjust-aggregation-tolerance} for details).

\begin{algorithm}[hbtp]
\caption{GVE-Louvain: Our multicore Louvain algorithm.}
\label{alg:louvain}
\begin{algorithmic}[1]
\Require{$G$: Input graph}
\Require{$C$: Community membership of each vertex}
\Require{$G'$: Input/super-vertex graph}
\Require{$C'$: Community membership of each vertex in $G'$}
\Require{$K'$: Total edge weight of each vertex}
\Require{$\Sigma'$: Total edge weight of each community}
\Ensure{$H_t$: Collision-free per-thread hashtable}
\Ensure{$l_i$: Number of iterations performed (per pass)}
\Ensure{$l_p$: Number of passes performed}
\Ensure{$\tau$: Per iteration tolerance}
\Ensure{$\tau_{agg}$: Aggregation tolerance}

\Statex

\Function{louvain}{$G$} \label{alg:louvain--begin}
  \State Vertex membership: $C \gets [0 .. |V|)$ \textbf{;} $G' \gets G$ \label{alg:louvain--initialization}
  \ForAll{$l_p \in [0 .. \text{\small{MAX\_PASSES}})$} \label{alg:louvain--passes-begin}
    \State $\Sigma' \gets K' \gets vertexWeights(G')$ \textbf{;} $C' \gets [0 .. |V'|)$ \label{alg:louvain--reset-weights}
    \State Mark all vertices in $G'$ as unprocessed \label{alg:louvain--reset-affected}
    \State $l_i \gets louvainMove(G', C', K', \Sigma')$ \Comment{Alg. \ref{alg:louvainlm}} \label{alg:louvain--local-move}
    \If{$l_i \le 1$} \textbf{break} \Comment{Globally converged?} \label{alg:louvain--globally-converged}
    \EndIf
    \State $|\Gamma|, |\Gamma_{old}| \gets$ Number of communities in $C$, $C'$
    \If{$|\Gamma|/|\Gamma_{old}| > \tau_{agg}$} \textbf{break} \Comment{Low shrink?} \label{alg:louvain--aggregation-tolerance}
    \EndIf
    \State $C' \gets$ Renumber communities in $C'$ \label{alg:louvain--renumber}
    \State $C \gets$ Lookup dendrogram using $C$ to $C'$ \label{alg:louvain--lookup}
    \State $G' \gets louvainAggregate(G', C')$ \Comment{Alg. \ref{alg:louvainag}} \label{alg:louvain--aggregate}
    \State $\tau \gets \tau / \text{\small{TOLERANCE\_DROP}}$ \Comment{Threshold scaling} \label{alg:louvain--threshold-scaling}
  \EndFor \label{alg:louvain--passes-end}
  \State $C \gets$ Lookup dendrogram using $C$ to $C'$ \label{alg:louvain--lookup-last}
  \Return{$C$} \label{alg:louvain--return}
\EndFunction \label{alg:louvain--end}
\end{algorithmic}
\end{algorithm}

If convergence is not achieved, we renumber the communities (line \ref{alg:louvain--renumber}), update the top-level community memberships $C$ using dendrogram lookup (line \ref{alg:louvain--lookup}), and proceed to the aggregation phase by calling \texttt{louvainAggregate()} (Algorithm \ref{alg:louvainag}). Additionally, we scale the convergence threshold for subsequent passes (line \ref{alg:louvain--threshold-scaling}) before continuing to the next pass (line \ref{alg:louvain--passes-begin}). Once all passes are completed, we perform a final update of the top-level community memberships $C$ using dendrogram lookup (line \ref{alg:louvain--lookup-last}) and return the top-level community membership $C$ for each vertex in $G$.

\subsubsection{Local-moving phase of GVE-Louvain}

The local-moving phase of GVE-Louvain, outlined in Algorithm \ref{alg:louvainlm}, iteratively moves vertices between communities to maximize modularity. This process is handled by the \texttt{louvainMove()} function, which takes as input the current graph $G'$, community membership $C'$, the total edge weight of each vertex $K'$, and the total edge weight of each community $\Sigma'$, and returns the number of iterations $l_i$ performed.

\begin{algorithm}[hbtp]
\caption{Local-moving phase of GVE-Louvain.}
\label{alg:louvainlm}
\begin{algorithmic}[1]
\Require{$G'$: Input/super-vertex graph}
\Require{$C'$: Community membership of each vertex}
\Require{$K'$: Total edge weight of each vertex}
\Require{$\Sigma'$: Total edge weight of each community}
\Ensure{$H_t$: Collision-free per-thread hashtable}
\Ensure{$l_i$: Number of iterations performed}
\Ensure{$\tau$: Per iteration tolerance}

\Statex

\Function{louvainMove}{$G', C', K', \Sigma'$} \label{alg:louvainlm--move-begin}
  \ForAll{$l_i \in [0 .. \text{\small{MAX\_ITERATIONS}})$} \label{alg:louvainlm--iterations-begin}
    \State Total delta-modularity per iteration: $\Delta Q \gets 0$ \label{alg:louvainlm--init-deltaq}
    \ForAll{unprocessed $i \in V'$ \textbf{in parallel}} \label{alg:louvainlm--loop-vertices-begin}
      \State Mark $i$ as processed (prune) \label{alg:louvainlm--prune}
      \State $H_t \gets scanCommunities(\{\}, G', C', i, false)$ \label{alg:louvainlm--scan}
      \State $\rhd$ Use $H_t, K', \Sigma'$ to choose best community
      \State $c^* \gets$ Best community linked to $i$ in $G'$ \label{alg:louvainlm--best-community-begin}
      \State $\delta Q^* \gets$ Delta-modularity of moving $i$ to $c^*$ \label{alg:louvainlm--best-community-end}
      \If{$c^* = C'[i]$} \textbf{continue} \label{alg:louvainlm--best-community-same}
      \EndIf
      \State $\Sigma'[C'[i]] -= K'[i]$ \textbf{;} $\Sigma'[c^*] += K'[i]$ \textbf{atomic} \label{alg:louvainlm--perform-move-begin}
      \State $C'[i] \gets c^*$ \textbf{;} $\Delta Q \gets \Delta Q + \delta Q^*$ \label{alg:louvainlm--perform-move-end}
      \State Mark neighbors of $i$ as unprocessed \label{alg:louvainlm--remark}
    \EndFor \label{alg:louvainlm--loop-vertices-end}
    \If{$\Delta Q \le \tau$} \textbf{break} \Comment{Locally converged?} \label{alg:louvainlm--locally-converged}
    \EndIf
  \EndFor \label{alg:louvainlm--iterations-end}
  \Return{$l_i$} \label{alg:louvainlm--return}
\EndFunction \label{alg:louvainlm--move-end}

\Statex

\Function{scanCommunities}{$H_t, G', C', i, self$}
  \ForAll{$(j, w) \in G'.edges(i)$}
    \If{\textbf{not} $self$ and $i = j$} \textbf{continue}
    \EndIf
    \State $H_t[C'[j]] \gets H_t[C'[j]] + w$
  \EndFor
  \Return{$H_t$}
\EndFunction
\end{algorithmic}
\end{algorithm}

\begin{algorithm}[hbtp]
\caption{Aggregation phase of GVE-Louvain.}
\label{alg:louvainag}
\begin{algorithmic}[1]
\Require{$G'$: Input/super-vertex graph}
\Require{$C'$: Community membership of each vertex}
\Ensure{$G'_{C'}$: Community vertices (CSR)}
\Ensure{$G''$: Super-vertex graph (weighted CSR)}
\Ensure{$*.offsets$: Offsets array of a CSR graph}
\Ensure{$H_t$: Collision-free per-thread hashtable}

\Statex

\Function{louvainAggregate}{$G', C'$}
  \State $\rhd$ Obtain vertices belonging to each community
  \State $G'_{C'}.offsets \gets countCommunityVertices(G', C')$ \label{alg:louvainag--coff-begin}
  \State $G'_{C'}.offsets \gets exclusiveScan(G'_{C'}.offsets)$ \label{alg:louvainag--coff-end}
  \ForAll{$i \in V'$ \textbf{in parallel}} \label{alg:louvainag--comv-begin}
    \State Add edge $(C'[i], i)$ to CSR $G'_{C'}$ atomically
  \EndFor \label{alg:louvainag--comv-end}
  \State $\rhd$ Obtain super-vertex graph
  \State $G''.offsets \gets communityTotalDegree(G', C')$ \label{alg:louvainag--yoff-begin}
  \State $G''.offsets \gets exclusiveScan(G''.offsets)$ \label{alg:louvainag--yoff-end}
  \State $|\Gamma| \gets$ Number of communities in $C'$
  \ForAll{$c \in [0, |\Gamma|)$ \textbf{in parallel}} \label{alg:louvainag--y-begin}
    \If{degree of $c$ in $G'_{C'} = 0$} \textbf{continue}
    \EndIf
    \State $H_t \gets \{\}$
    \ForAll{$i \in G'_{C'}.edges(c)$}
      \State $H_t \gets scanCommunities(H, G', C', i, true)$
    \EndFor
    \ForAll{$(d, w) \in H_t$}
      \State Add edge $(c, d, w)$ to CSR $G''$ atomically
    \EndFor
  \EndFor \label{alg:louvainag--y-end}
  \Return $G''$ \label{alg:louvainag--return}
\EndFunction
\end{algorithmic}
\end{algorithm}

The main loop of the local-moving phase is presented in lines \ref{alg:louvainlm--iterations-begin}-\ref{alg:louvainlm--iterations-end}. Initially, the total delta-modularity per iteration, $\Delta Q$, is initialized in line \ref{alg:louvainlm--init-deltaq}. Next, we iterate over unprocessed vertices in parallel (lines \ref{alg:louvainlm--loop-vertices-end}-\ref{alg:louvainlm--loop-vertices-end}). For each vertex $i$, we mark it as processed (vertex pruning, line \ref{alg:louvainlm--prune}), scan the communities connected to it (excluding itself) in line \ref{alg:louvainlm--scan}. The best community $c*$ for moving $i$ to is then determined (line \ref{alg:louvainlm--best-community-begin}), and the delta-modularity of this move is calculated (line \ref{alg:louvainlm--best-community-end}). If a better community is found, we update $i$'s community membership (lines \ref{alg:louvainlm--perform-move-begin}-\ref{alg:louvainlm--perform-move-end}) and mark its neighbors as unprocessed (line \ref{alg:louvainlm--remark}). Convergence of the local-moving phase is checked in line \ref{alg:louvainlm--locally-converged}, and the loop breaks if convergence is achieved or if \textit{\small{MAX\_ITERATIONS}} is reached. Finally,\ignore{in line \ref{alg:louvainlm--return},} we return the number of iterations performed, $l_i$.

\subsubsection{Aggregation phase of GVE-Louvain}

Finally, we present the pseudocode for the aggregation phase in Algorithm \ref{alg:louvainag}, where we aggregate communities into super-vertices. Here, the \texttt{louvainAggre} \texttt{gate()} function processes the current graph $G'$ and the community membership $C'$ as inputs, and returns the super-vertex graph $G''$.

In lines \ref{alg:louvainag--coff-begin}-\ref{alg:louvainag--coff-end}, we construct the offsets array $G'_{C'}.offsets$ for the community vertices CSR by first counting the vertices in each community using \texttt{countCommunityVertices()} and then performing an exclusive scan\ignore{on the resulting array}. Next, in lines \ref{alg:louvainag--comv-begin}-\ref{alg:louvainag--comv-end}, we iterate over all vertices in parallel, atomically populating the\ignore{community vertices} CSR $G'_{C'}$ with vertices assigned to their respective communities. To build the offsets array for the super-vertex graph CSR, we over-estimate the degree of each super-vertex by calculating the total degree of each community using \texttt{communityTotalDegree()} and perform an exclusive scan\ignore{on the array} (lines \ref{alg:louvainag--yoff-begin}-\ref{alg:louvainag--yoff-end}). This results in a \textit{holey} CSR, with gaps in the edges and weights arrays. In lines \ref{alg:louvainag--y-begin}-\ref{alg:louvainag--y-end}, we process all communities $c \in [0, |\Gamma|)$ in parallel, using \texttt{scanCommunities()} (defined in Algorithm \ref{alg:louvainlm}) to identify and store all linked communities $d$ (with edge weights $w$) for each vertex $i$ in community $c$ into a per-thread hashtable $H_t$. Once $H_t$ contains all linked communities and their weights for community $c$, we atomically add them as edges to super-vertex $c$ in the super-vertex graph $G''$. Finally, in line \ref{alg:louvainag--return}, we return the constructed super-vertex graph $G''$.

\subsection{Optimizations for GPU-based Louvain\ignore{algorithm}}

Our GPU-based Louvain builds on top of GVE-Louvain and incorporates: \textbf{(1)} An asynchronous parallel algorithm with a single community membership vector, allowing threads work independently on different parts of the graph --- promoting faster convergence; \textbf{(2)} A limit of $20$ iterations per pass and $10$ passes; \textbf{(3)} Threshold scaling optimization using an initial tolerance $\tau = 0.01$ and a tolerance drop rate of $10$, \textbf{(4)} An aggregation tolerance $\tau_{agg}$ of $0.8$ for avoiding aggregation when only a small fraction of communities are to be hierarchically merged; \textbf{(5)} Vertex pruning to reduce redundant computations, where vertices mark their neighbors as unprocessed upon label change; and \textbf{(6)} Use of parallel prefix sum and preallocated CSR data structures for identifying community vertices and storing the intermediate super-vertex graphs. We now discuss our GPU-specific techniques for the Louvain method, below.

\subsubsection{Mitigating community swaps}

We first observe that the GPU implementation of Louvain algorithm, like LPA \cite{sahu2024nulpa}, fails to converge for certain input graphs. This issue arises when interconnected vertices are trapped in cycles of community swaps, repeatedly moving to each other's community. This can when the vertices are symmetrically connected to each other’s communities. Such swaps are more common than expected, as GPUs execute in lockstep, and symmetric vertices can swap community memberships if assigned to the same SM. Since SM assignment is based on vertex IDs, this issue can persist indefinitely, preventing convergence. 

\begin{figure}[hbtp]
  \centering
  \subfigure{
    \label{fig:voptswapprevent--runtime}
    \includegraphics[width=0.98\linewidth]{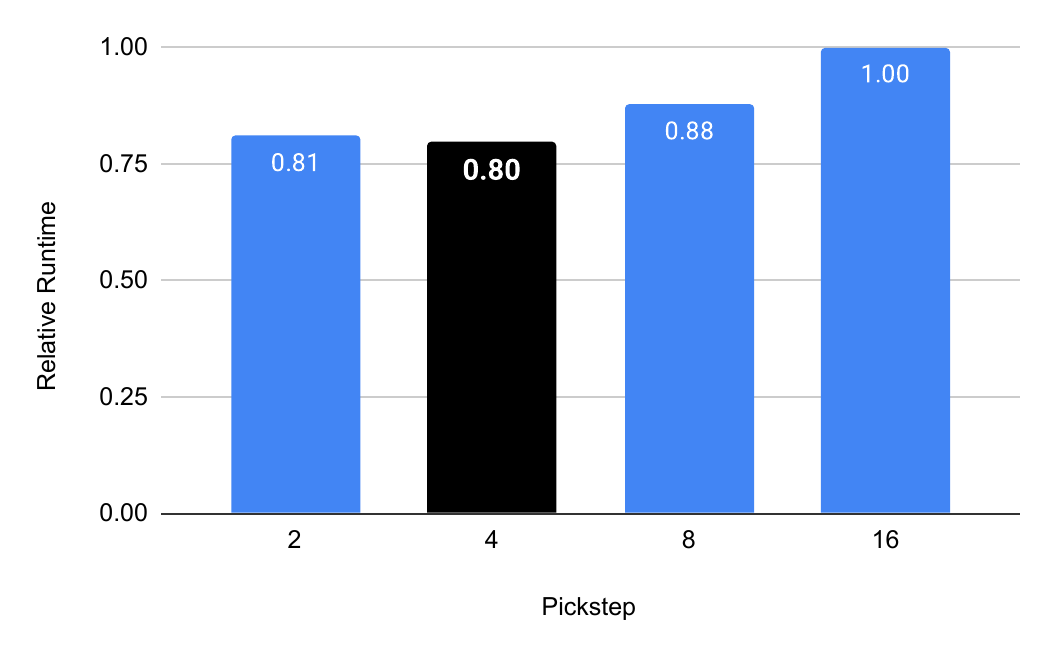}
  }
  \subfigure{
    \label{fig:voptswapprevent--modularity}
    \includegraphics[width=0.98\linewidth]{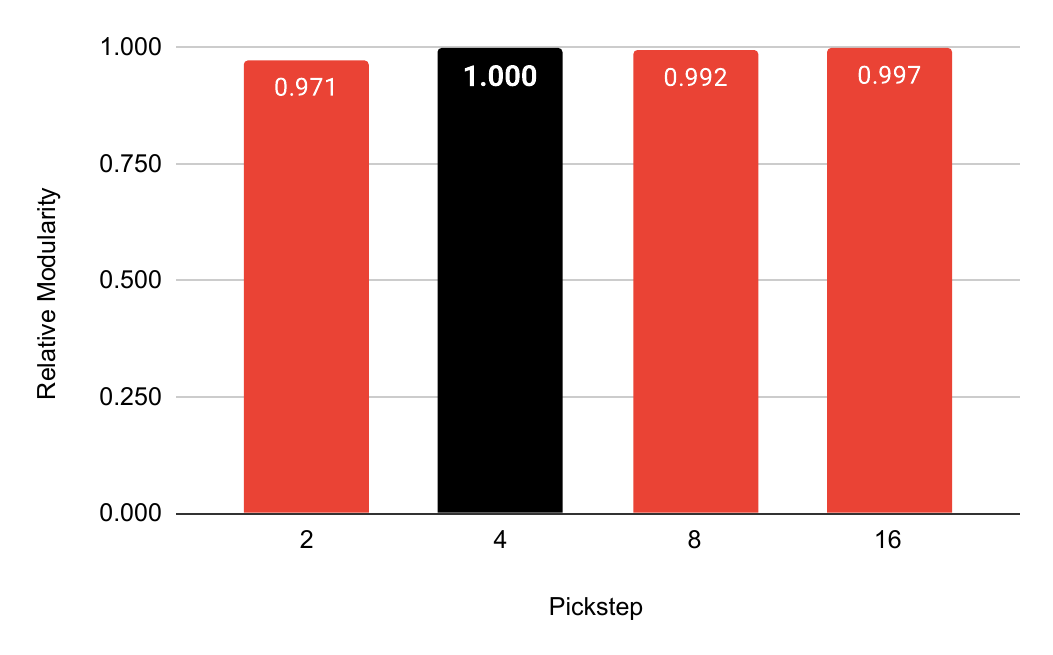}
  } \\[-2ex]
  \caption{Relative Runtime and Modularity of the communities obtained using the \textit{Pick-Less (PL)} community swap prevention technique, which restricts label selection to those with lower ID values every $2$ to $16$ iterations.}
  \label{fig:voptswapprevent}
\end{figure}

To address this challenge, we explore the \textbf{Pick-Less (PL)} method, where a vertex can only switch to a new community if its new community ID is lower than the current one. This ensures that one vertex involved in a swap does not change its community, thus breaking the symmetry. However, applying this method too frequently can limit the algorithm's ability to find high-quality communities. We experiment with applying the PL method every $2$, $4$, $8$, or $16$ iterations of the Louvain algorithm, on large graphs from Table \ref{tab:dataset}. Figures \ref{fig:voptswapprevent--runtime} and \ref{fig:voptswapprevent--modularity} show the mean relative runtime and modularity for each method. As the figures show, applying PL every \textbf{4 iterations} (PL4) yields the highest modularity, while being $1.25\times$ faster than PL16. Thus, we adopt the PL4 method for our GPU implementation of the Louvain algorithm.

\subsubsection{Hashtable design}

In GVE-Louvain, we utilized per-thread collision-free hashtables, with each hashtable consisting of a key list and a full-size values array (size $|V|$), kept separated in memory. It yielded a $4.4\times$ performance improvement over C++'s \texttt{std::unorde} \texttt{red\_map}. However, GPUs operate with far more threads, making it impractical to allocate $O(|V|)$ memory per thread. This issue is further complicated by the limited memory capacity of GPUs.

\begin{figure}[hbtp]
  \centering
  \subfigure{
    \label{fig:vabout-hashtable--all}
    \includegraphics[width=0.86\linewidth]{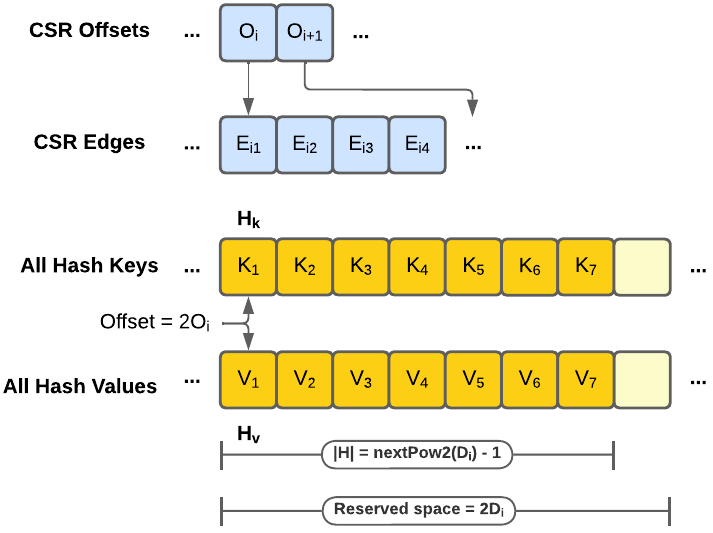}
  } \\[-2ex]
  \caption{Per-vertex open-addressing hashtables are used in our GPU implementation of Louvain. Each vertex $i$ is assigned a hashtable $H$, which consists of a keys array $H_k$ and a values array $H_v$. Memory for the hash key and value arrays of all vertices is allocated as a contiguous block. The offset of vertex $i$'s hashtable within this memory block is $2O_i$, where $O_i$ is the vertex's CSR offset. The memory allocated for each hashtable is $2D_i$, where $D_i$ represents the vertex's degree. The hashtable's capacity, defined as the maximum number of key-value pairs it can store, is $\text{nextPow2}(D_i) - 1$ \cite{sahu2024nulpa}.}
  \label{fig:vabout-hashtable}
\end{figure}

To overcome this, we reserve a dedicated hashtable of size $2D_i$ for each vertex $i$ in the graph, where $D_i$ is its degree, ensuring the load factor remains below $100\%$ and bounding the total memory usage by the hashtables to $O(|E|)$. These hashtables use open addressing and consist of two arrays: a \textit{keys array} $H_k$ for neighboring labels and a \textit{values array} $H_v$ for edge weights. Setting up the hashtable requires only two memory allocations of size $2|E|$, one for $H_k$ and one for $H_v$. Each thread retrieves the hashtable for a vertex $i$ using information from the input graph's CSR, specifically the starting position of its edges and its degree. For efficient hash computation, the hashtable size is set to $nextPow2(D_i) - 1$, allowing the use of the $\bmod$ operator as a simple hash function. Figure \ref{fig:vabout-hashtable} illustrates this per-vertex hashtable design. The hashtable supports key operations such as accumulating weights for matching keys, clearing entries, and identifying the key with the highest weight. Thread safety during access and updates is ensured using atomic operations.

\begin{figure}[hbtp]
  \centering
  \subfigure{
    \label{fig:voptresolution--runtime}
    \includegraphics[width=0.98\linewidth]{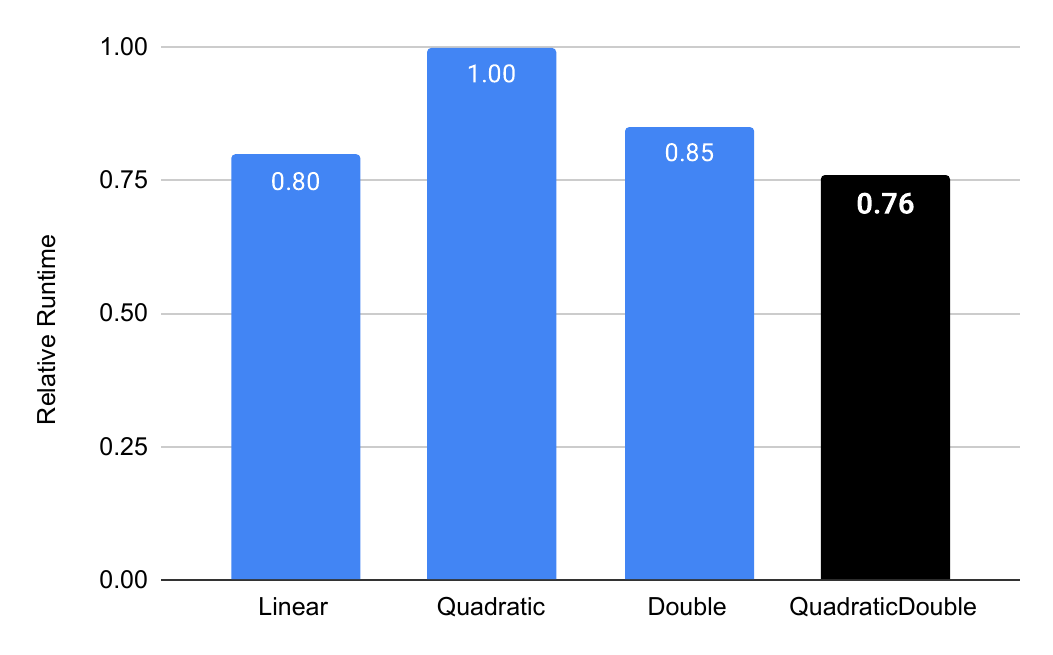}
  } \\[-2ex]
  \caption{Relative runtime when using \textit{Linear} probing, \textit{Quadratic} probing, \textit{Double} hashing, and a hybrid of \textit{Quadratic} probing and \textit{Double} hashing (\textit{Quadratic-double}) for collision resolution in per-vertex hash tables --- used for storing the total associated weight with each neighboring community, or the delta-modularity of moving to each community.}
  \label{fig:voptresolution}
\end{figure}

We now consider collision resolution techniques for use with the hashtables. \textit{Linear probing} is cache-efficient but suffers from high clustering, which increases the likelihood of repeated collisions. \textit{Double hashing}, on the other hand, significantly reduces clustering but is less cache-efficient. \textit{Quadratic probing} offers a middle ground, balancing clustering and cache efficiency. In our implementation, we set the probe step for linear probing to $1$. For quadratic probing, we start with an initial probe step of $1$ and double it after each collision. For double hashing, we use a secondary prime, $p_2 = \text{nextPow2}(p_1) - 1$, which is co-prime with $p_1 = |H|$, and calculate the probe step using the modulo operator. However, repeated collisions can cause thread divergence, significantly impacting performance. To further minimize collisions, we also examine a hybrid \textit{quadratic-double} method, which combines quadratic probing and double hashing. Figure \ref{fig:voptresolution} presents the mean relative runtime of our GPU-based Louvain using these four collision resolution techniques (with per-vertex hashtables), on large graphs from Table \ref{tab:dataset}.

As shown in Figure \ref{fig:voptresolution}, the \textbf{quadratic-double} approach achieves the best performance, being $1.05\times$, $1.32\times$, and $1.12\times$ faster than Louvain implementations using linear probing, quadratic probing, and double hashing, respectively. This indicates that the hybrid approach optimally balances clustering and cache efficiency, leading us to adopt it for our GPU-based Louvain implementation. We also tested shared memory-based hashtables for low-degree vertices, but it yielded negligible performance gains.

\subsubsection{Selecting datatype for hashtable values}

\begin{figure}[hbtp]
  \centering
  \subfigure{
    \label{fig:voptdatatype--runtime}
    \includegraphics[width=0.98\linewidth]{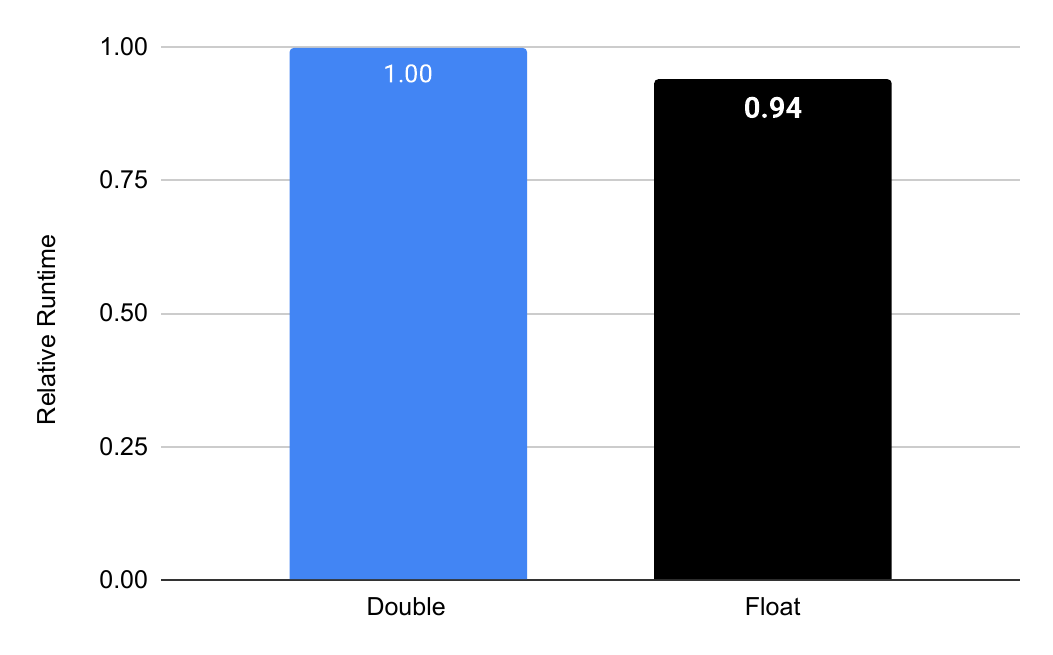}
  } \\[-2ex]
  \caption{Relative Runtime when using 32-bit floating-point values (\textit{Float}) instead of 64-bit floating-point values (\textit{Double}) for hashtable values. Note that weighted degrees of vertices $K$, total edge weights of communities $\Sigma$, and all computations remain in 64-bit floating-point precision.}
  \label{fig:voptdatatype}
\end{figure}

Next, we evaluate the use of 32-bit floating-point numbers instead of 64-bit for aggregated label weights in hashtables. As shown in Figure \ref{fig:voptdatatype}, this change maintains community quality while offering a moderate speedup. Consequently, we adopt 32-bit floats for $H_v$ in our GPU implementation of the Louvain algorithm.

\begin{figure}[hbtp]
  \centering
  \subfigure{
    \label{fig:voptsdegmove--runtime}
    \includegraphics[width=0.98\linewidth]{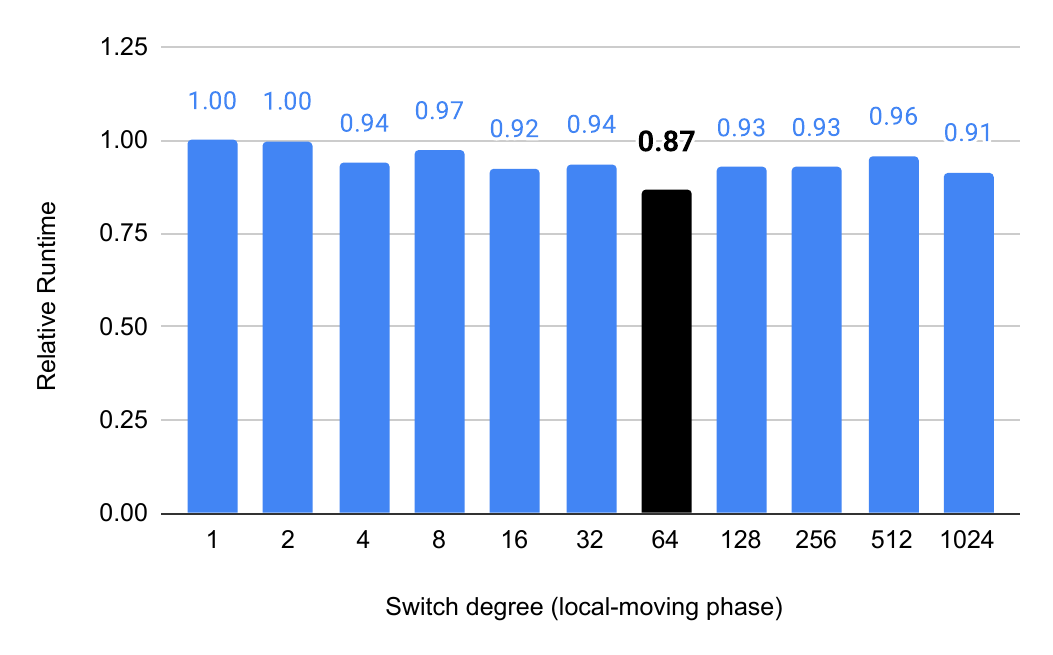}
  } \\[-2ex]
  % \subfigure{
  %   \label{fig:voptsdegmove--modularity}
  %   \includegraphics[width=0.98\linewidth]{out/voptsdegmove-modularity.pdf}
  % } \\[-2ex]
  \caption{Relative runtime with varying switch degrees for the local-moving phase, where a degree-based condition in the thread- and block-per-vertex kernels decides whether a vertex is processed by a specific kernel.}
  \label{fig:voptsdegmove}
\end{figure}

\begin{figure}[hbtp]
  \centering
  \subfigure{
    \label{fig:voptsdegaggr--runtime}
    \includegraphics[width=0.98\linewidth]{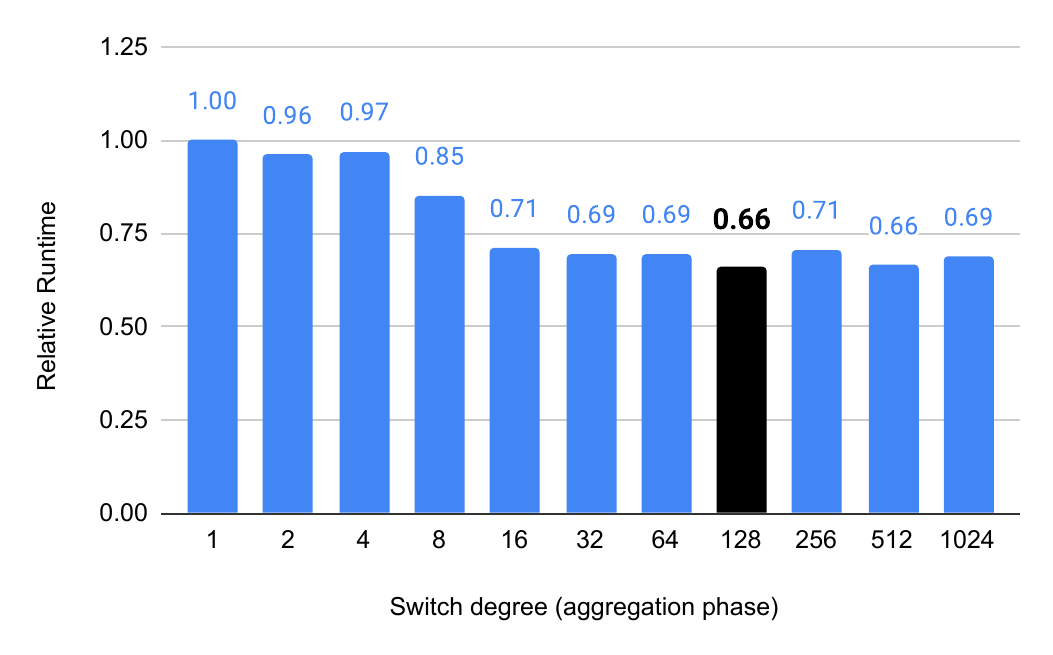}
  } \\[-2ex]
  % \subfigure{
  %   \label{fig:voptsdegaggr--modularity}
  %   \includegraphics[width=0.98\linewidth]{out/voptsdegaggr-modularity.pdf}
  % } \\[-2ex]
  \caption{Relative runtime with varying switch degrees for the aggregation phase, where a degree-based condition in the thread- and block-per-vertex kernels decides whether a vertex is processed by a specific kernel.}
  \label{fig:voptsdegaggr}
\end{figure}

\subsubsection{Partitioning work between two kernels}

Processing each vertex with a thread-block per vertex is inefficient for low-degree vertices since threads remain idle when the degree is below $32$, the warp size on NVIDIA GPUs. Given the prevalence of such vertices, we adopt a thread-per-vertex approach how low-degree vertices \cite{wu2010efficient}. Unlike $\nu$-LPA \cite{sahu2024nulpa} however, we cannot partition vertices for separate processing due to the random ordering in the super-vertex graph after each pass. Instead, we use a degree-based condition within both thread- and block-per-vertex kernels to determine execution. These kernels run over all vertices in both the input and super-vertex graphs. To determine an optimal switch degree, we vary it from $1$ to $1024$, evaluate performance across graphs in Table \ref{tab:dataset}, and plot mean relative runtime. Results in Figures \ref{fig:voptsdegmove} and \ref{fig:voptsdegaggr} indicate that a switch degree of $64$ is suitable for the local-moving phase, while a switch degree of $128$ is suitable for the aggregation phase. We also tested a fine-grained per-thread approach for scanning communities to reduce thread divergence, but it yielded no performance gains --- likely due to additional thread divergence offsetting benefits from reduced collisions in the hashtable.

\subsection{Our optimized GPU-based Louvain implementation}

The psuedocode of our GPU implementation of Louvain, which we refer to as $\nu$-Louvain, is given in Algorithms \ref{alg:vlouvain}, \ref{alg:vlouvainlm}, and \ref{alg:vlouvainag}. In $\nu$-Louvain, the symbol $\nu$ is a reference to ``video card". Similar to GVE-Louvain, $\nu$-Louvain has a time complexity of $O(KM)$, where $K$ is the number of iterations performed, and a space complexity of $O(M)$. In contrast, GVE-Louvain has a space complexity of $O(TN + M)$, where $T$ represents the number of threads being used.

\section{Evaluation}
\label{sec:evaluation}
\subsection{Experimental Setup}
\label{sec:setup}

\subsubsection{System used}
\label{sec:system}

We use two servers for our evaluations. The first features two Intel Xeon Gold 6226R processors (16 cores each, $2.90$ GHz) with $1$ MB L1 cache per core, $16$ MB L2 cache, and $22$ MB shared L3 cache, along with $512$ GB of RAM, running CentOS Stream 8. For GPU-based evaluations, we use a server equipped with an NVIDIA A100 GPU (108 SMs, 64 CUDA cores per SM, $80$ GB global memory, $1935$ GB/s bandwidth, and $164$ KB shared memory per SM), powered by an AMD EPYC-7742 processor (64 cores, $2.25$ GHz) with $512$ GB DDR4 RAM, running Ubuntu 20.04.

\subsubsection{Configuration}
\label{sec:configuration}

We use 32-bit integers for vertex identifiers and 32-bit floating-point numbers for edge weights and hash table values, but use 64-bit floats everything else. Compilation on the CPU-only system is done with GCC 8.5 and OpenMP 4.5, while the GPU-based system uses GCC 9.4, OpenMP 5.0, and CUDA 11.4.

\subsubsection{Dataset}
\label{sec:dataset}

The graphs used in our experiments, listed in Table \ref{tab:dataset}, are sourced from the SuiteSparse Matrix Collection \cite{suite19} and include web graphs, social networks, road networks, and protein k-mer graphs. The vertices in these graphs range from $3.07$ million to $214$ million and edges range from $25.4$ million to $3.80$ billion. All edges are undirected and weighted, with a default weight of $1$.

\begin{table}[hbtp]
  \centering
  \caption{List of $13$ graphs obtained SuiteSparse Matrix Collection \cite{suite19} (directed graphs are marked with $*$). Here, $|V|$ is the number of vertices, $|E|$ is the number of edges (after adding reverse edges), and $D_{avg}$ is the average degree, and $|\Gamma|$ is the number of communities obtained with \textit{GVE-Louvain}.}
  \label{tab:dataset}
  \begin{tabular}{|c||c|c|c|c|}
    \toprule
    \textbf{Graph} &
    \textbf{\textbf{$|V|$}} &
    \textbf{\textbf{$|E|$}} &
    \textbf{\textbf{$D_{avg}$}} &
    \textbf{\textbf{$|\Gamma|$}} \\
    % \textbf{$1 - \Gamma_G$} \\
    \midrule
    \multicolumn{5}{|c|}{\textbf{Web Graphs (LAW)}} \\ \hline
    indochina-2004$^*$ & 7.41M & 341M & 41.0 & 4.24K \\ \hline  % & \num{4.7e-4} & 2.9 GB
    uk-2002$^*$ & 18.5M & 567M & 16.1 & 42.8K \\ \hline  % & \num{9.6e-5} & 16 GB
    arabic-2005$^*$ & 22.7M & 1.21B & 28.2 & 3.66K \\ \hline  % & \num{5.5e-4} & 11 GB
    uk-2005$^*$ & 39.5M & 1.73B & 23.7 & 20.8K \\ \hline  % & \num{9.6e-5} & 16 GB
    webbase-2001$^*$ & 118M & 1.89B & 8.6 & 2.76M \\ \hline  % & \num{7.3e-7} & 18 GB
    it-2004$^*$ & 41.3M & 2.19B & 27.9 & 5.28K \\ \hline  % & \num{3.8e-4} & 19 GB
    sk-2005$^*$ & 50.6M & 3.80B & 38.5 & 3.47K \\ \hline  % & \num{5.8e-4} & 33 GB
    \multicolumn{5}{|c|}{\textbf{Social Networks (SNAP)}} \\ \hline
    com-LiveJournal & 4.00M & 69.4M & 17.4 & 2.54K \\ \hline  % & \num{7.9e-4} & 480 MB
    com-Orkut & 3.07M & 234M & 76.2 & 29 \\ \hline  % & \num{6.7e-2} & 1.7 GB
    \multicolumn{5}{|c|}{\textbf{Road Networks (DIMACS10)}} \\ \hline
    asia\_osm & 12.0M & 25.4M & 2.1 & 2.38K \\ \hline  % & \num{8.4e-4} & 200 MB
    europe\_osm & 50.9M & 108M & 2.1 & 3.05K \\ \hline  % & \num{6.6e-4} & 910 MB
    \multicolumn{5}{|c|}{\textbf{Protein k-mer Graphs (GenBank)}} \\ \hline
    kmer\_A2a & 171M & 361M & 2.1 & 21.2K \\ \hline  % & \num{9.4e-5} & 3.2 GB
    kmer\_V1r & 214M & 465M & 2.2 & 6.17K \\ \hline  % & \num{3.2e-4} & 4.2 GB
  \bottomrule
  \end{tabular}
\end{table}

\begin{figure*}[hbtp]
  \centering
  \subfigure[Runtime in seconds (logarithmic scale) with \textit{Vite (Louvain)}, \textit{Grappolo (Louvain)}, \textit{NetworKit Louvain}, \textit{cuGraph Louvain}, and \textit{GVE-Louvain}]{
    \label{fig:louvain-compare--runtime}
    \includegraphics[width=0.98\linewidth]{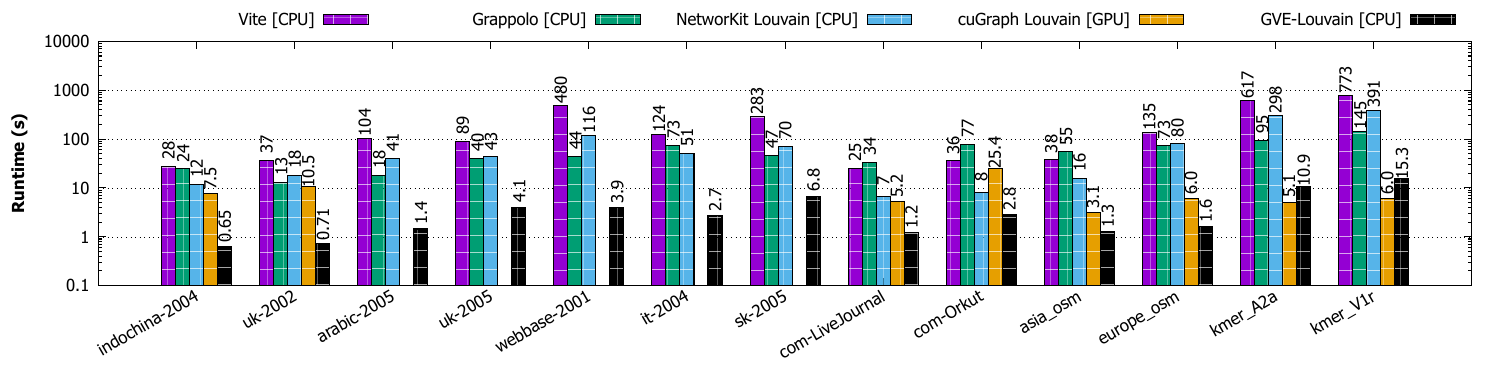}
  } \\[-0ex]
  \subfigure[Speedup of \textit{GVE-Louvain} with respect to \textit{Vite (Louvain)}, \textit{Grappolo (Louvain)}, \textit{NetworKit Louvain}, and \textit{cuGraph Louvain}.]{
    \label{fig:louvain-compare--speedup}
    \includegraphics[width=0.98\linewidth]{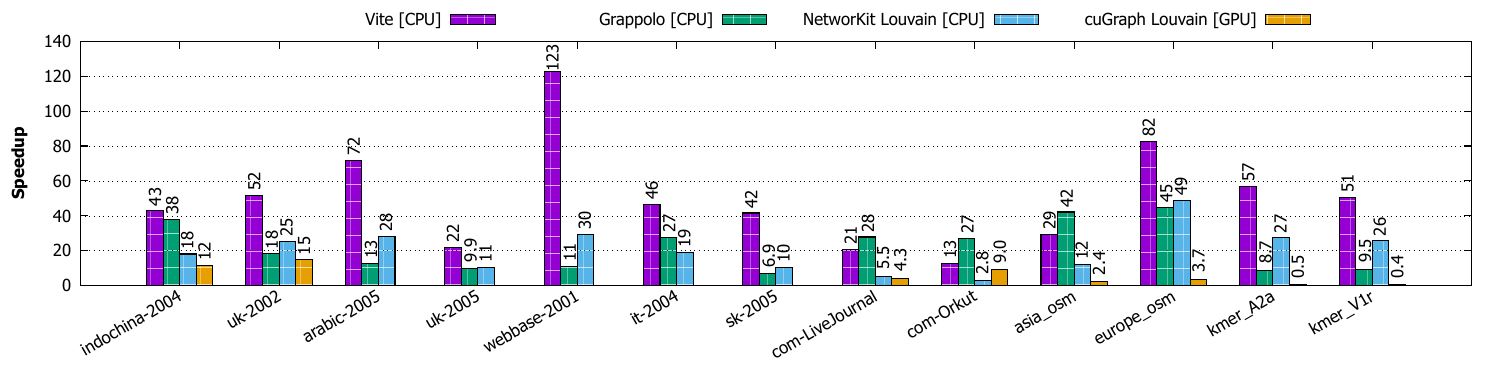}
  } \\[-0ex]
  \subfigure[Modularity of communities obtained with \textit{Vite (Louvain)}, \textit{Grappolo (Louvain)}, \textit{NetworKit Louvain}, \textit{cuGraph Louvain}, and \textit{GVE-Louvain}.]{
    \label{fig:louvain-compare--modularity}
    \includegraphics[width=0.98\linewidth]{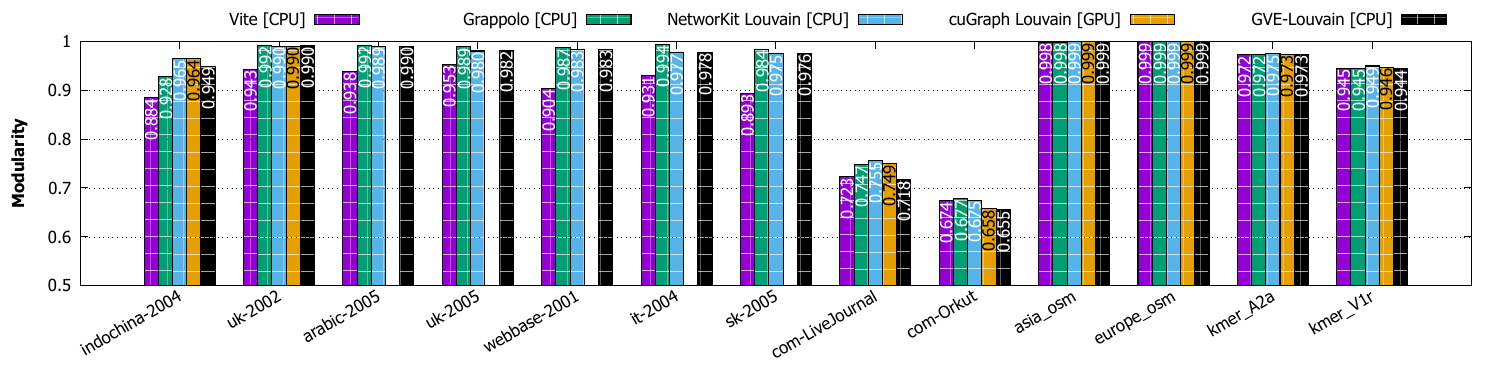}
  } \\[-2ex]
  \caption{Runtime in seconds (logarithmic scale), speedup, and modularity of communities obtained with \textit{Vite (Louvain)}, \textit{Grappolo (Louvain)}, \textit{NetworKit Louvain}, \textit{cuGraph Louvain}, and \textit{GVE-Louvain} for each graph in the dataset.}
  \label{fig:louvain-compare}
\end{figure*}

\subsection{Performance Comparison}
\label{sec:comparison}

\subsubsection{Comparing Performance of GVE-Louvain}

We now compare the performance of GVE-Louvain with Vite (Louvain) \cite{ghosh2018scalable}, Grappolo (Louvain) \cite{com-halappanavar17}, NetworKit Louvain \cite{staudt2016networkit}, and cuGraph Louvain \cite{kang2023cugraph}. Vite, Grappolo, and NetworKit are CPU-based multicore implementations, while cuGraph Louvain is a GPU-based implementation. For Vite, we convert the graph datasets to its binary format, run it on a single node with threshold cycling/scaling optimization, and measure the reported average total time. Grappolo is executed on the same system, and we record its reported total runtime. For NetworKit Louvain, we use a Python script to invoke \texttt{PLM} (Parallel Louvain Method) and retrieve the total time via \texttt{getTiming()}. To test cuGraph Louvain, we write a Python script that configures the Rapids Memory Manager (RMM) with a pool allocator for fast memory allocations, then execute \texttt{cugraph.louvain()} on the loaded graph. For each implementation, we measure both the runtime and the modularity of the resulting communities with each graph, averaging the results over five runs. When using cuGraph, we exclude the runtime of the first run to ensure that subsequent measurements accurately reflect RMM's pool usage without the overhead of initial CUDA memory allocation.

Figure \ref{fig:louvain-compare--runtime} shows the runtimes of Vite (Louvain), Grappolo (Louvain), NetworKit Louvain, cuGraph Louvain, and GVE-Louvain across the dataset. We observe that cuGraph's Louvain algorithm fails to run on the \textit{arabic-2005}, \textit{uk-2005}, \textit{webbase-2001}, \textit{it-2004}, and \textit{sk-2005} graphs due to out-of-memory issues. On the \textit{sk-2005} graph, GVE-Louvain processes communities in $6.8$ seconds, achieving a rate of $560$ million edges/s. Figure \ref{fig:louvain-compare--speedup} highlights the speedup of GVE-Louvain compared to the other implementations, showing it is on average $50\times$, $22\times$, $20\times$, and $3.2\times$ faster than Vite, Grappolo, NetworKit Louvain, and cuGraph Louvain, respectively. Additionally, Figure \ref{fig:louvain-compare--modularity} compares the modularity of communities found by each implementation. On average, GVE-Louvain achieves $3.1\%$ higher modularity than Vite (particularly on web graphs), $0.6\%$ lower modularity than Grappolo and NetworKit (especially on poorly clustered social networks), and $0.7\%$ lower modularity than cuGraph Louvain\ignore{(for graphs where cuGraph Louvain runs)}.

\subsubsection{Comparing Performance of $\nu$-Louvain}

Next, we evaluate the performance of $\nu$-Louvain by comparing it with state-of-the-art multicore and GPU-based implementations: Grappolo (Louvain) \cite{com-halappanavar17}, NetworKit Louvain \cite{staudt2016networkit}, Nido (Louvain) \cite{sahu2023gvelouvain}, and cuGraph Louvain \cite{kang2023cugraph}. Grappolo and NetworKit are multicore CPU implementations, while Nido and cuGraph Louvain run on GPUs. For Nido, we convert graph datasets to binary format, similar to Vite, and execute it on a single A100 GPU with luby coloring enabled.\ignore{The other implementations follow their standard execution methods, as described above.} As earlier, for each graph we measure runtime and modularity (as reported) over five runs to compute the average.

\begin{figure*}[hbtp]
  \centering
  \subfigure[Runtime in seconds (logarithmic scale) with \textit{Grappolo (Louvain)}, \textit{NetworKit Louvain}, \textit{Nido (Louvain)}, \textit{cuGraph Louvain}, and \textit{$\nu$-Louvain}]{
    \label{fig:vcompare--runtime}
    \includegraphics[width=0.98\linewidth]{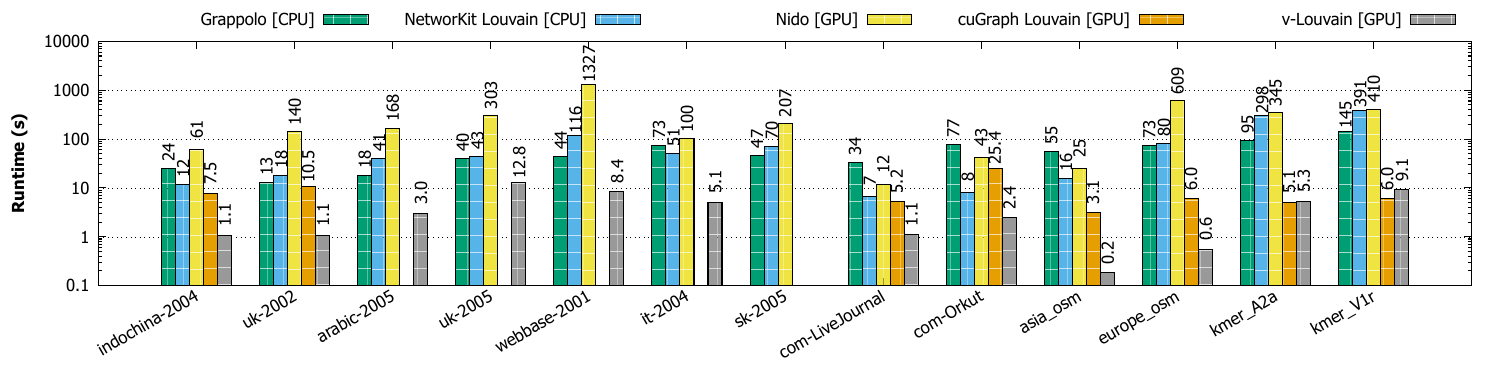}
  }
  \subfigure[Speedup of \textit{$\nu$-Louvain} (logarithmic scale) with respect to \textit{Grappolo (Louvain)}, \textit{NetworKit Louvain}, \textit{Nido (Louvain)}, and \textit{cuGraph Louvain}.]{
    \label{fig:vcompare--speedup}
    \includegraphics[width=0.98\linewidth]{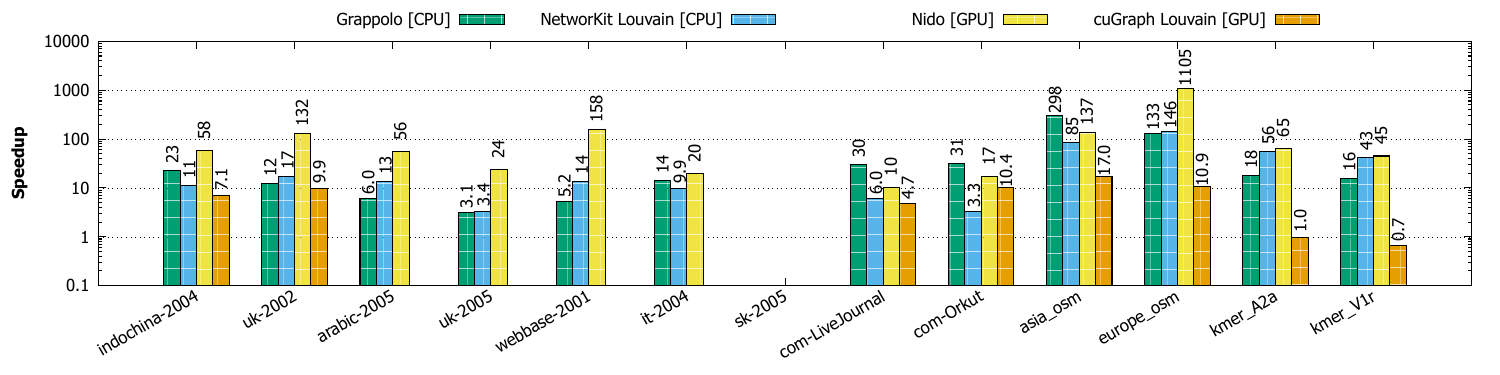}
  }
  \subfigure[Modularity of communities obtained with \textit{Grappolo (Louvain)}, \textit{NetworKit Louvain}, \textit{Nido (Louvain)}, \textit{cuGraph Louvain}, and \textit{$\nu$-Louvain}.]{
    \label{fig:vcompare--modularity}
    \includegraphics[width=0.98\linewidth]{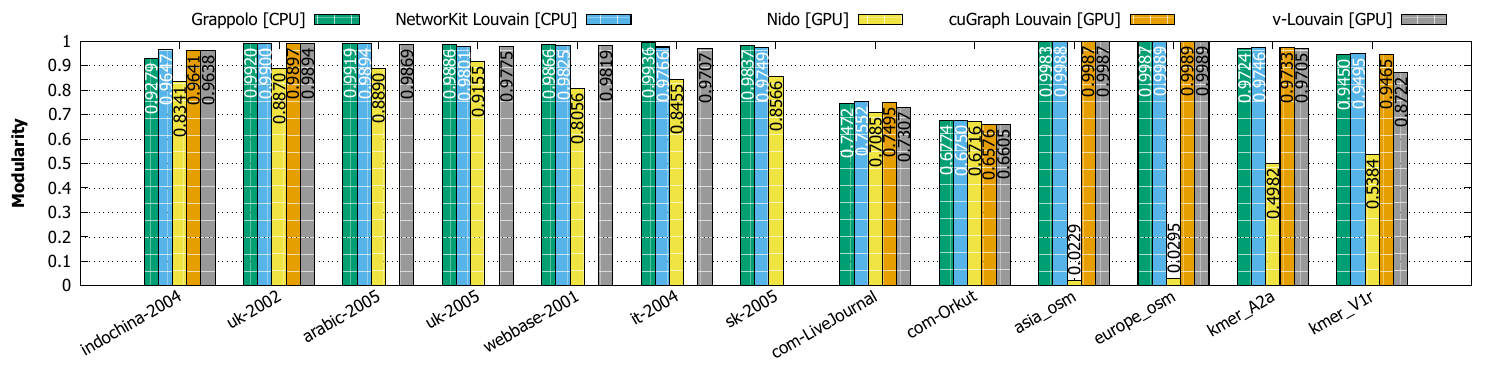}
  } \\[-2ex]
  \caption{Runtime in seconds (log-scale), speedup (log-scale), and modularity of obtained communities with \textit{Grappolo (Louvain)}, \textit{NetworKit Louvain}, \textit{Nido (Louvain)}, \textit{cuGraph Louvain}, and \textit{$\nu$-Louvain} for each graph in the dataset.}
  \label{fig:vcompare}
\end{figure*}

\begin{figure*}[hbtp]
  \centering
  \subfigure[Runtime in seconds (logarithmic scale) with \textit{$\nu$-Louvain} and \textit{GVE-Louvain}]{
    \label{fig:vpick--runtime}
    \includegraphics[width=0.98\linewidth]{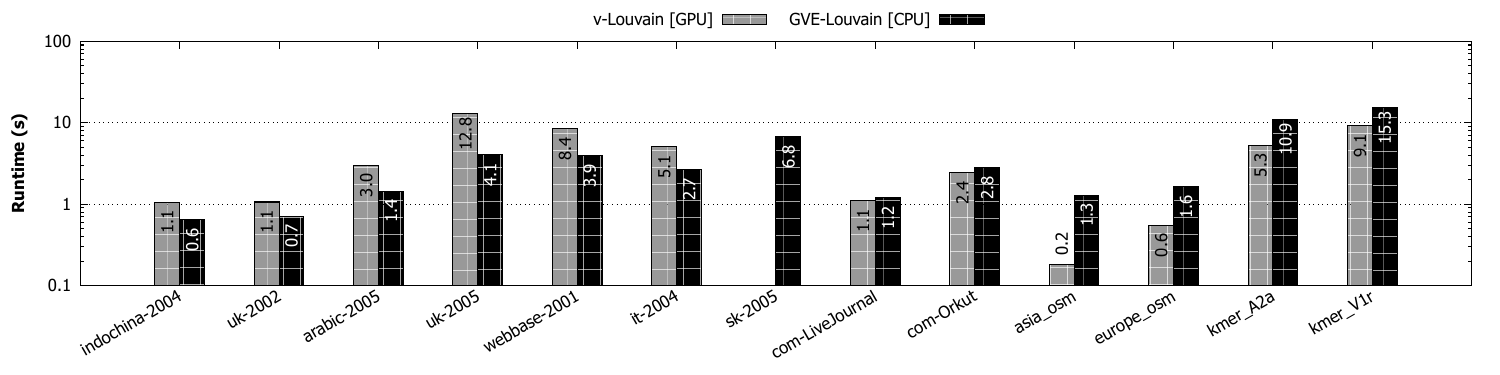}
  } \\[-0ex]
  \subfigure[Speedup of \textit{$\nu$-Louvain} (logarithmic scale) with respect to \textit{GVE-Louvain}.]{
    \label{fig:vpick--speedup}
    \includegraphics[width=0.98\linewidth]{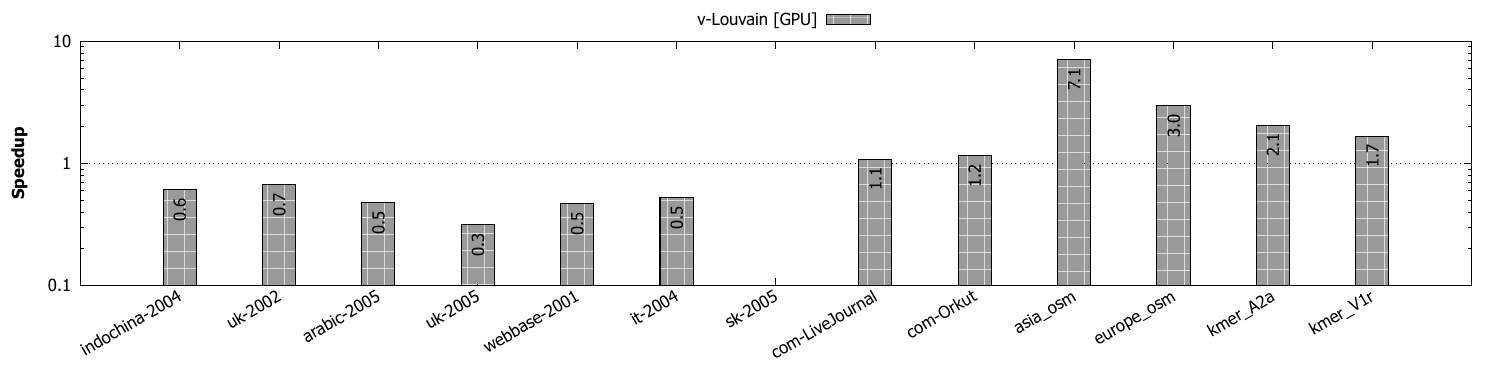}
  } \\[-0ex]
  \subfigure[Modularity of communities obtained with \textit{$\nu$-Louvain} and \textit{GVE-Louvain}.]{
    \label{fig:vpick--modularity}
    \includegraphics[width=0.98\linewidth]{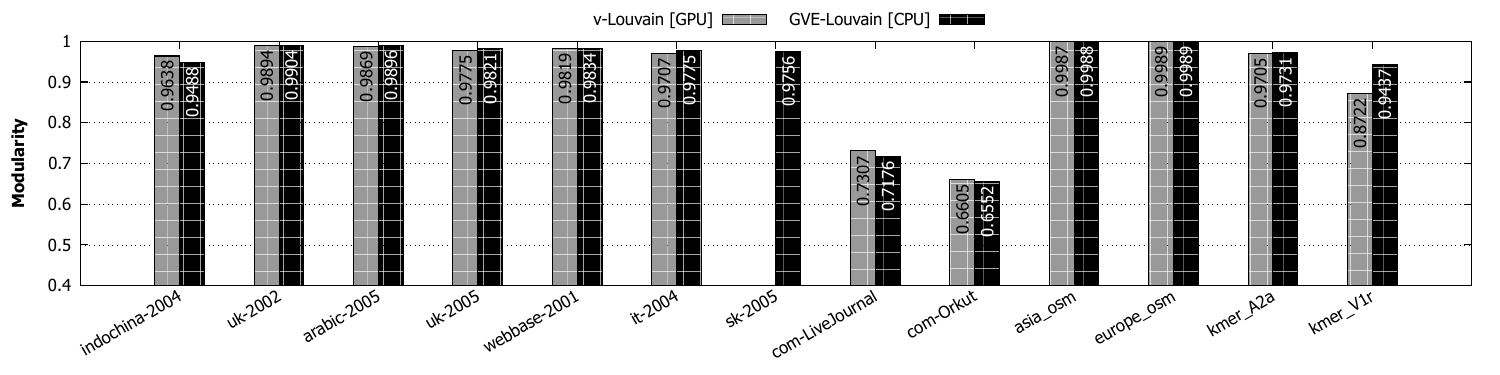}
  } \\[-2ex]
  \caption{Runtime in seconds (log-scale), speedup (log-scale), and modularity of communities obtained with \textit{$\nu$-Louvain} and \textit{GVE-Louvain} for each graph in the dataset.}
  \label{fig:vpick}
\end{figure*}

Figure \ref{fig:vcompare--runtime} presents the runtimes of Grappolo, NetworKit Louvain, Nido, cuGraph Louvain, and $\nu$-Louvain for each graph. Figure \ref{fig:vcompare--speedup} shows the speedup of $\nu$-Louvain compared to the other implementations. cuGraph Louvain fails on the \textit{arabic-2005}, \textit{uk-2005}, \textit{webbase-2001}, \textit{it-2004}, and \textit{sk-2005} graphs, while $\nu$-Louvain fails on the \textit{sk-2005} graph due to out-of-memory issues, so these results are excluded. $\nu$-Louvain achieves average speedups of $20\times$, $17\times$, $61\times$, and $5.0\times$ over Grappolo, NetworKit Louvain, Nido, and cuGraph Louvain, respectively. On the \textit{it-2004} graph, $\nu$-Louvain identifies communities in $5.1$ seconds, achieving a processing rate of $405$ million edges/s. Figure \ref{fig:vcompare--modularity} shows the modularity of communities identified by each implementation. On average, $\nu$-Louvain has $1.1\%$, $1.2\%$, and $1.3\%$ lower modularity than Grappolo, NetworKit Louvain, and cuGraph Louvain (where cuGraph Louvain runs), but $45\%$ higher modularity than Nido. We also note that GVE-Louvain is, on average, $56\times$ faster than Nido, while obtaining communities of $43\%$ higher modularity. $\nu$-Louvain yields a lower modularity on the \textit{kmer\_V1r} graph, but due to the asynchronous nature of the algorithm, higher modularity may be achieved in different runs\ignore{, potentially matching the other Louvain implementations}. 

\begin{figure*}[hbtp]
  \centering
  \subfigure[Phase split]{
    \label{fig:louvain-splits--phase}
    \includegraphics[width=0.47\linewidth]{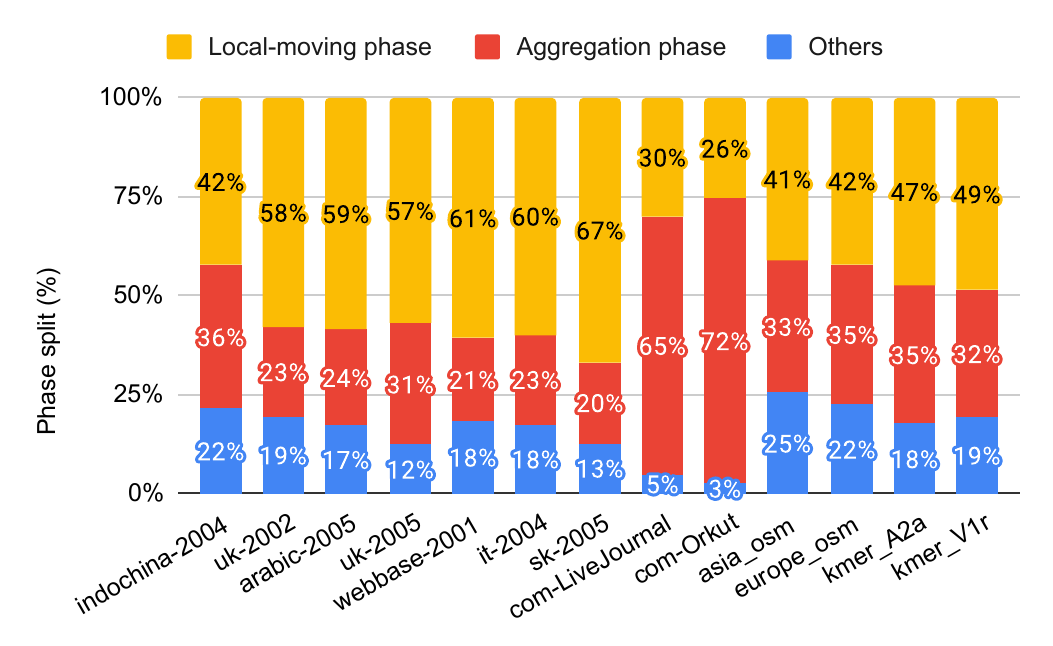}
  }
  \subfigure[Pass split]{
    \label{fig:louvain-splits--pass}
    \includegraphics[width=0.47\linewidth]{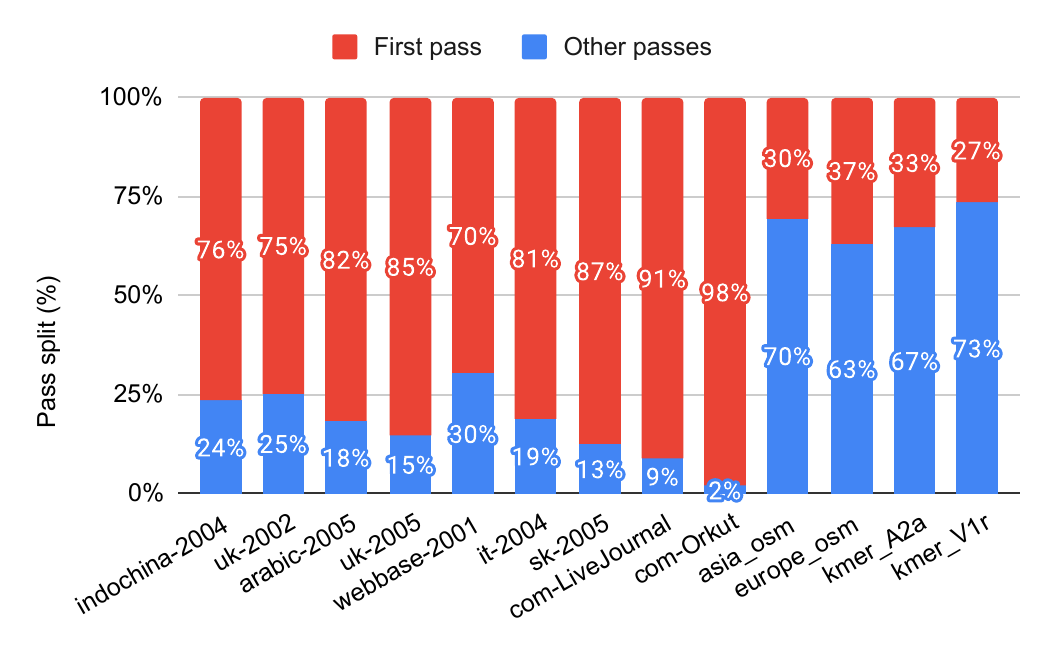}
  } \\[-2ex]
  \caption{Phase split of \textit{GVE-Louvain} shown on the left, and pass split shown on the right for each graph in the dataset. A pass consists of the local-moving and the aggregation phases, and multiple passes of the algorithm are run, until convergence.}
  \label{fig:louvain-splits}
\end{figure*}

\begin{figure}[hbtp]
  \centering
  \subfigure{
    \label{fig:louvain-hardness--all}
    \includegraphics[width=0.98\linewidth]{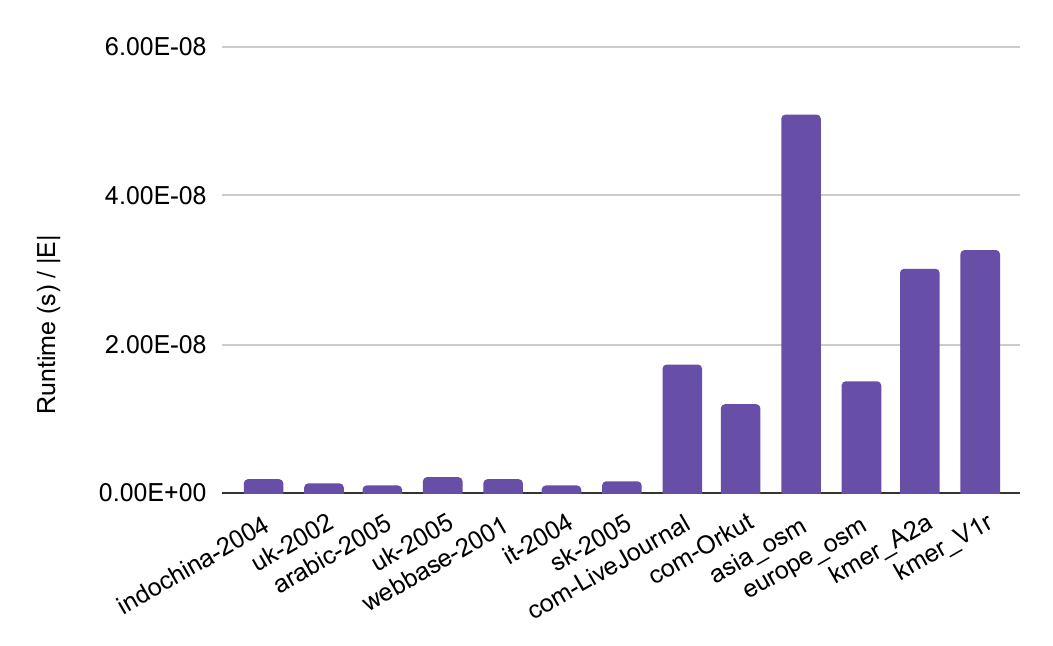}
  } \\[-2ex]
  \caption{Runtime $/ |E|$ factor with \textit{GVE-Louvain} for each graph in the dataset.}
  \label{fig:louvain-hardness}
\end{figure}

\begin{figure}[hbtp]
  \centering
  \includegraphics[width=0.98\linewidth]{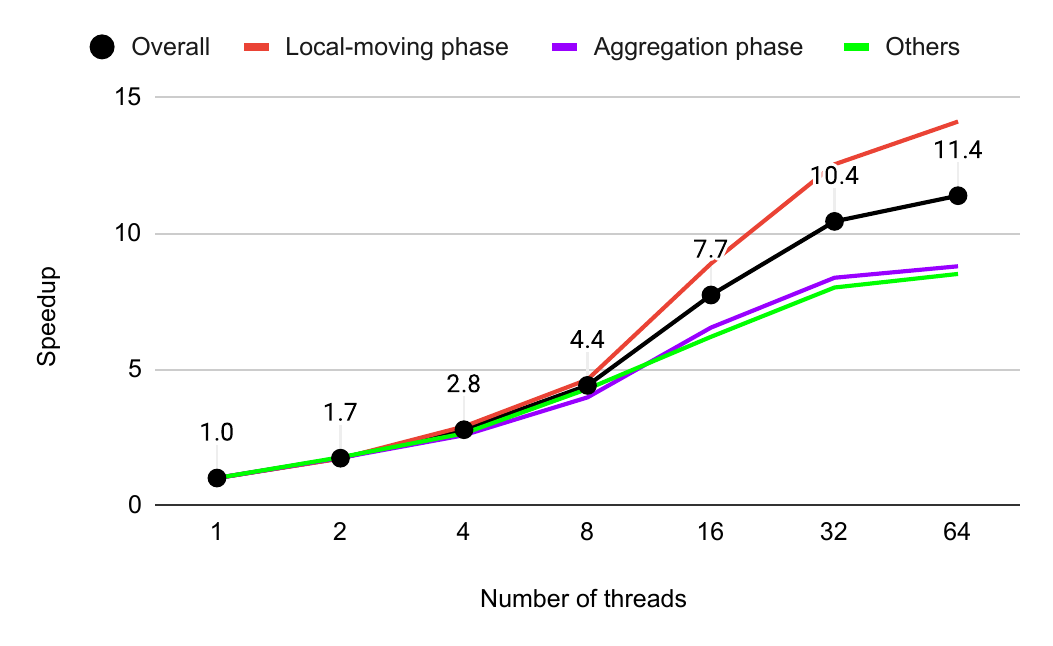} \\[-2ex]
  \caption{Overall speedup of \textit{GVE-Louvain}, and its various phases (local-moving, aggregation, others), with increasing number of threads (in multiples of 2).}
  \label{fig:louvain-ss}
\end{figure}

\subsubsection{Comparing Performance of GVE-Louvain with $\nu$-Louvain}

We now compare the performance of GVE-Louvain and $\nu$-Louvain. Figure \ref{fig:vpick--runtime} shows their runtimes, and Figure \ref{fig:vpick--speedup} illustrates the speedup of $\nu$-Louvain over GVE-Louvain. $\nu$-Louvain encounters out-of-memory issues on the \textit{sk-2005} graph, so those results are excluded. On average, $\nu$-Louvain achieves only a $1.03\times$ speedup over GVE-Louvain --- but is particularly faster on road networks. Figure \ref{fig:vpick--modularity} compares the modularity of the communities identified, where $\nu$-Louvain has $0.5\%$ lower modularity than GVE-Louvain.

Thus, running on a A100 GPU, $\nu$-Louvain, does not necessarily outperform GVE-Louvain. This lack of performance improvement likely stems from the reduced parallelism in the later passes of the algorithm, where only a (relatively) small number of super-vertices are being processed. While using the GPU for the first pass and CPUs for the remaining passes is an option, the overhead of managing both devices likely negates any potential runtime and energy efficiency gains. Additionally, the limited memory capacity of GPUs restricts the size of graphs that can be processed. In contrast, multicore implementations of the Louvain method are simpler and offer strong performance. Therefore, we expect that CPU architectures, with their higher clock speeds and greater efficiency for serial or mixed workloads, may be better suited for community detection. These findings are expected to extend to the Leiden algorithm \cite{com-traag19}.

\subsection{Performance Analysis}

\subsubsection{Analyzing Performance of GVE-Louvain}

Figures \ref{fig:louvain-splits--phase} and \ref{fig:louvain-splits--pass} illustrate the phase-wise and pass-wise breakdowns of GVE-Louvain’s runtime. The phase-wise breakdown shows that GVE-Louvain spends most of its time in the local-moving phase for \textit{web graphs}, \textit{road networks}, and \textit{protein k-mer graphs}, while it spends more time in the aggregation phase for \textit{social networks}. The pass-wise breakdown reveals that the first pass dominates the runtime on high-degree graphs (such as \textit{web graphs} and \textit{social networks}), while subsequent passes take longer on low-degree graphs (\textit{road networks} and \textit{protein k-mer graphs}). On average, $49\%$ of GVE-Louvain's runtime is spent in the local-moving phase, $35\%$ in the aggregation phase, and $16\%$ in other steps (including initialization, renumbering communities, looking up the dendrogram, and resetting communities). Additionally, $67\%$ of the runtime occurs in the first pass, which is the most expensive due to the size of the original graph, with later passes operating on super-vertex graphs \cite{com-wickramaarachchi14}.
We also observe that graphs with lower average degree (such as \textit{road networks} and \textit{protein k-mer graphs}) and those with poor community structure (e.g., \verb|com-LiveJournal| and \verb|com-Orkut|) tend to exhibit a higher $\text{runtime}/|E|$ ratio, as shown in Figure \ref{fig:louvain-hardness}.

\begin{figure*}[hbtp]
  \centering
  \subfigure[Phase split]{
    \label{fig:vsplit--phase}
    \includegraphics[width=0.48\linewidth]{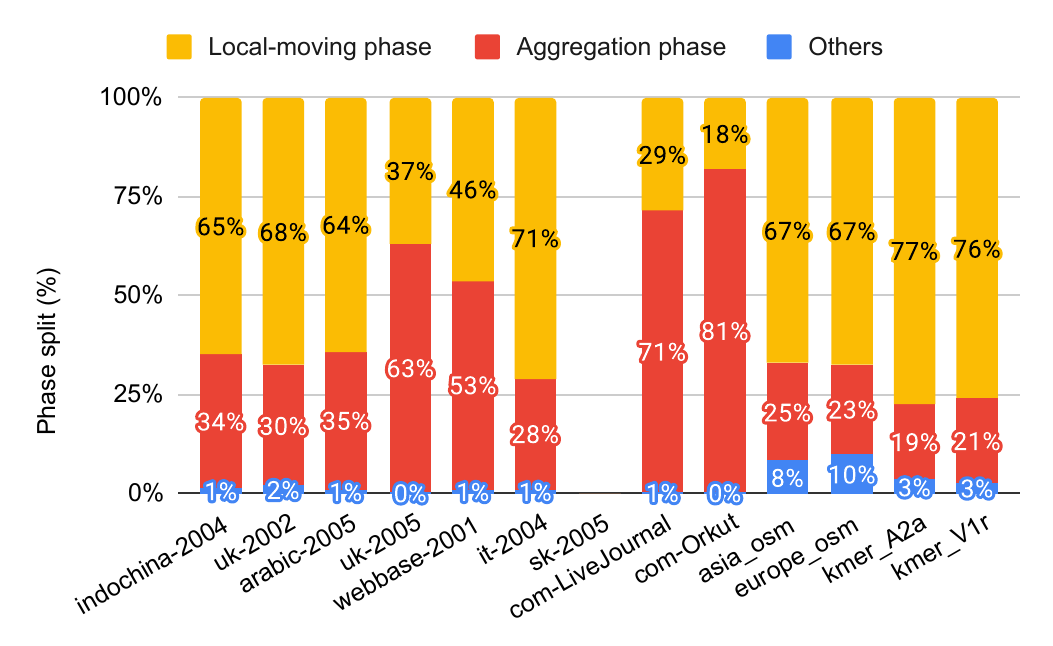}
  }
  \subfigure[Pass split]{
    \label{fig:vsplit--pass}
    \includegraphics[width=0.48\linewidth]{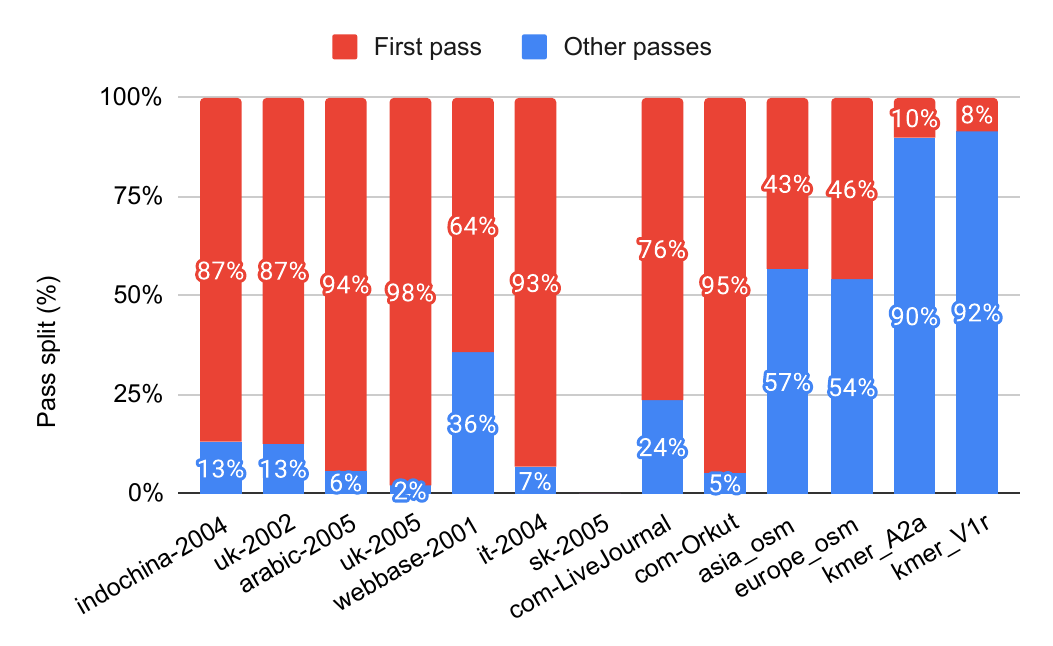}
  } \\[-2ex]
  \caption{Phase split and pass split of $\nu$-Louvain. Note that a pass consists of the local-moving and the aggregation phases, and multiple passes of the algorithm are run, until convergence.}
  \label{fig:vsplit}
\end{figure*}

\subsubsection{Analyzing Performance of $\nu$-Louvain}

Figures \ref{fig:vsplit--phase} and \ref{fig:vsplit--pass} show the phase-wise and pass-wise splits of $\nu$-Louvain. As Figure \ref{fig:vsplit--phase} shows, the aggregation phase dominates runtime, except for social networks and two web graphs. On high-degree graphs (web and social networks), the first pass takes the most time, while later passes dominate low-degree graphs (road networks and protein k-mer graphs) (Figure \ref{fig:vsplit--pass}). On average, $57\%$ of the runtime is spent in the local-moving phase, $40\%$ in aggregation, and $3\%$ on other steps (initialization, renumbering, dendrogram lookup, and resetting). Further, $67\%$ of the runtime is spent on the first pass.

\subsection{Strong Scaling of GVE-Louvain}

Finally, we evaluate the strong scaling performance of GVE-Louvain by varying the number of threads from 1 to 64 in powers of 2 for each input graph. For each configuration, we measure the total execution time for community detection, along with its phase splits (local-moving, aggregation, and others), averaging the results over five runs. As shown in Figure \ref{fig:louvain-ss}, GVE-Louvain achieves an average speedup of $10.4\times$ with 32 threads compared to a single-threaded execution. This corresponds to a $1.6\times$ speedup per thread doubling. Scaling is constrained by sequential phases in the algorithm. At 64 threads, both NUMA and hyper-threading further limit performance (our server has 32 cores), yielding a speedup of only $11.4\times$.

\section{Conclusion}
\label{sec:conclusion}
In conclusion, in this report we presented GVE-Louvain, our parallel multicore implementation of the Louvain algorithm, that, to our knowledge, is the most efficient implementation on multicore CPUs. On a dual 16-core Intel Xeon Gold 6226R server, GVE-Louvain achieves speedups of $50\times$, $22\times$, $20\times$, and $3.2\times$ over Vite (Multicore), Grappolo (Multicore), NetworKit Louvain (Multicore), and cuGraph Louvain (on an NVIDIA A100 GPU), respectively. On a web graph with $3.8B$ edges, it identifies communities in $6.8$ seconds, achieving a processing rate of $560M$ edges/s and a strong scaling factor of $1.6\times$ per thread doubling. We also introduced $\nu$-Louvain, our efficient GPU-based implementation, and evaluated it against Grappolo, NetworKit Louvain, Nido (on an NVIDIA A100 GPU), and cuGraph Louvain, where it achieved speedups of $20\times$, $17\times$, $61\times$, and $5.0\times$, respectively. However, it only performed on par with GVE-Louvain. This lack of performance improvement mainly stems from reduced workload and parallelism in later algorithm passes, a limitation likely affecting the Leiden algorithm as well. These results suggest that multicore CPUs, with higher clock speeds and superior handling of irregular workloads, are better suited for community detection than GPUs running more complex algorithms.

%% The acknowledgments section.
\begin{acks}
I would like to thank Prof. Kishore Kothapalli, Prof. Dip Sankar Banerjee, and Souvik Karfa for their support.
\end{acks}

%% Bibliography style to be used, and the bibliography file.
\bibliographystyle{ACM-Reference-Format}
\bibliography{main}

\clearpage
\appendix
\section{Appendix}

\subsubsection{Main step of our GPU-based Louvain}

The main function of $\nu$-Louvain is \texttt{louvain()}, which is given in Algorithm \ref{alg:vlouvain}. It includes initialization, local-moving phase, and aggregation phase. The function takes as input a graph $G$, and outputs the community membership $C$ for each vertex.

Initially, the community membership $C$ is set for all vertices, such that each vertex is its own community (line \ref{alg:vlouvain--initialization}). Next, we perform up to $MAX\_PASSES$ iterations (lines \ref{alg:vlouvain--passes-begin}-\ref{alg:vlouvain--passes-end}) of the Louvain algorithm, or until convergence in the community memberships is achieved. In each pass, the total edge weight per vertex $K'$, the total edge weight per community $\Sigma'$, and the community membership $C'$ for vertices in the current graph $G'$ are initialized (line \ref{alg:vlouvain--reset-weights}), and all vertices are marked as unprocessed (line \ref{alg:vlouvain--reset-affected}). Next in line \ref{alg:vlouvain--local-move}, the local-moving phase is performed through the \texttt{louvainMove()} function (see Algorithm \ref{alg:vlouvainlm}), which optimizes community assignments. If the local-moving phase converged in a single iteration (line \ref{alg:vlouvain--globally-converged}), or if the community count $|\Gamma|$ drop is minimal (line \ref{alg:vlouvain--aggregation-tolerance}), we terminate early since global convergence is reached. If convergence has not been achieved, communities are renumbered (line \ref{alg:vlouvain--renumber}), top-level memberships $C$ are updated using dendrogram lookup (line \ref{alg:vlouvain--lookup}), \texttt{louvainAggregate()} (Algorithm \ref{alg:vlouvainag}) is invoked to perform the aggregation phase, and the convergence threshold is scaled for subsequent passes (line \ref{alg:vlouvain--threshold-scaling}). This process repeats until convergence, after which a final update of $C$ is performed via dendrogram lookup (line \ref{alg:vlouvain--lookup-last}), and $C$ is returned (line \ref{alg:vlouvain--return}).

\begin{algorithm}[hbtp]
\caption{$\nu$-Louvain: Our GPU-based Louvain algorithm.}
\label{alg:vlouvain}
\begin{algorithmic}[1]
\Require{$G$: Input graph}
\Require{$C$: Community membership of each vertex}
\Require{$G'$: Input/super-vertex graph}
\Require{$C'$: Community membership of each vertex in $G'$}
\Require{$K'$: Total edge weight of each vertex}
\Require{$\Sigma'$: Total edge weight of each community}
\Ensure{$l_i$: Number of iterations performed (per pass)}
\Ensure{$l_p$: Number of passes performed}
\Ensure{$\tau$: Per iteration tolerance}
\Ensure{$\tau_{agg}$: Aggregation tolerance}

\Statex

\Function{$\nu$louvain}{$G$} \label{alg:vlouvain--begin}
  \State Vertex membership: $C \gets [0 .. |V|)$ \textbf{;} $G' \gets G$ \label{alg:vlouvain--initialization}
  \ForAll{$l_p \in [0 .. \text{\small{MAX\_PASSES}})$} \label{alg:vlouvain--passes-begin}
    \State $\Sigma' \gets K' \gets vertexWeights(G')$ \textbf{;} $C' \gets [0 .. |V'|)$ \label{alg:vlouvain--reset-weights}
    \State Mark all vertices in $G'$ as unprocessed \label{alg:vlouvain--reset-affected}
    \State $l_i \gets \nu louvainMove(G', C', K', \Sigma')$ \Comment{Alg. \ref{alg:vlouvainlm}} \label{alg:vlouvain--local-move}
    \If{$l_i \le 1$} \textbf{break} \Comment{Globally converged?} \label{alg:vlouvain--globally-converged}
    \EndIf
    \State $|\Gamma|, |\Gamma_{old}| \gets$ Number of communities in $C$, $C'$
    \If{$|\Gamma|/|\Gamma_{old}| > \tau_{agg}$} \textbf{break} \Comment{Low shrink?} \label{alg:vlouvain--aggregation-tolerance}
    \EndIf
    \State $C' \gets$ Renumber communities in $C'$ \label{alg:vlouvain--renumber}
    \State $C \gets$ Lookup dendrogram using $C$ to $C'$ \label{alg:vlouvain--lookup}
    \State $G' \gets \nu louvainAggregate(G', C')$ \Comment{Alg. \ref{alg:vlouvainag}} \label{alg:vlouvain--aggregate}
    \State $\tau \gets \tau / \text{\small{TOLERANCE\_DROP}}$ \Comment{Threshold scaling} \label{alg:vlouvain--threshold-scaling}
  \EndFor \label{alg:vlouvain--passes-end}
  \State $C \gets$ Lookup dendrogram using $C$ to $C'$ \label{alg:vlouvain--lookup-last}
  \Return{$C$} \label{alg:vlouvain--return}
\EndFunction \label{alg:vlouvain--end}
\end{algorithmic}
\end{algorithm}

\subsubsection{Local-moving phase of $\nu$-Louvain}

The pseudocode for the local-moving phase of $\nu$-Louvain is detailed in Algorithm \ref{alg:vlouvainlm}. It iteratively moves vertices between communities to maximize modularity. The primary function, \texttt{louvainMove()}, takes as input a super-vertex graph $G'$, the community membership $C'$ of each vertex, the total edge weight $K'$ of each vertex, and the total edge weight $\Sigma'$ of each community. It returns the number of iterations $l_i$ performed in the local-moving phase.

\begin{algorithm}[hbtp]
\caption{Local-moving phase of $\nu$-Louvain.}
\label{alg:vlouvainlm}
\begin{algorithmic}[1]
\Require{$G'$: Input/super-vertex graph}
\Require{$C'$: Community membership of each vertex}
\Require{$K'$: Total edge weight of each vertex}
\Require{$\Sigma'$: Total edge weight of each community}
\Ensure{$H$: Per-vertex hashtable ($H_k$: keys, $H_v$: values)}
\Ensure{$l_i$: Number of iterations performed}
\Ensure{$\tau$: Per iteration tolerance}

\Statex

\Function{$\nu$louvainMove}{$G', C', K', \Sigma'$} \label{alg:vlouvainlm--move-begin}
  \ForAll{$l_i \in [0 .. \text{\small{MAX\_ITERATIONS}})$} \label{alg:vlouvainlm--iterations-begin}
    \State $\rhd$ Mitigate community swaps with \textbf{pick-less} mode
    \If{$(l_i + \rho/2) \bmod \rho = 0$} Enable \textbf{pick-less} mode
    \Else\ Disable \textbf{pick-less} mode
    \EndIf
    \State Total delta-modularity per iteration: $\Delta Q \gets 0$ \label{alg:vlouvainlm--init-deltaq}
    \ForAll{unprocessed $i \in V'$ \textbf{in parallel}} \label{alg:vlouvainlm--loop-vertices-begin}
      \State Mark $i$ as processed (prune) \label{alg:vlouvainlm--prune}
      \State $\rhd$ If degree of $i < \text{\small{SWITCH\_DEGREE\_MOVE}}$,
      \State $\rhd$ process using a thread, else use a thread-block
      \State $p_1 \gets nextPow2(G'.degree(i)) - 1$
      \State $p_2 \gets nextPow2(p_1) - 1$
      \State $\theta_H \gets 2 * G'.offset(i)$
      \State $H_k \gets buf_k[\theta_H : \theta_H + p_1]$ \Comment{$H$ is \textbf{shared}, if using}
      \State $H_v \gets buf_v[\theta_H : \theta_H + p_1]$ \Comment{a thread-block}
      \State $hashtableClear(H)$ \textbf{in parallel}
      \ForAll{$(j, w) \in G'.neighbors(i)$ \textbf{in parallel}}
        \If{$j = i$} \textbf{continue}
        \EndIf
        \State $hashtableAccumulate(H, p_1, p_2, C[j], w)$ \label{alg:vlouvainlm--scan}
      \EndFor
      \State $\rhd$ Use $H, K', \Sigma'$ to choose best community
      \State $c^* \gets$ Best community linked to $i$ in $G'$ \label{alg:vlouvainlm--best-community-begin}
      \State $\delta Q^* \gets$ Delta-modularity of moving $i$ to $c^*$ \label{alg:vlouvainlm--best-community-end}
      \If{$c^* = C'[i]$} \textbf{continue} \label{alg:vlouvainlm--best-community-same}
      \EndIf
      \If{is \textbf{pick-less and} $c^* > C'[i]$} \textbf{continue}
      \EndIf
      \State $\Sigma'[C'[i]] -= K'[i]$ \textbf{;} $\Sigma'[c^*] += K'[i]$ \textbf{atomic} \label{alg:vlouvainlm--perform-move-begin}
      \State $C'[i] \gets c^*$ \textbf{;} $\Delta Q \gets \Delta Q + \delta Q^*$ \label{alg:vlouvainlm--perform-move-end}
      \State Mark neighbors of $i$ as unprocessed \label{alg:vlouvainlm--remark}
    \EndFor \label{alg:vlouvainlm--loop-vertices-end}
    \If{$\Delta Q \le \tau$} \textbf{break} \Comment{Locally converged?} \label{alg:vlouvainlm--locally-converged}
    \EndIf
  \EndFor \label{alg:vlouvainlm--iterations-end}
  \Return{$l_i$} \label{alg:vlouvainlm--return}
\EndFunction \label{alg:vlouvainlm--move-end}

\Statex

\Function{scanCommunities}{$H_t, G', C', i, self$}
  \ForAll{$(j, w) \in G'.edges(i)$}
    \If{\textbf{not} $self$ and $i = j$} \textbf{continue}
    \EndIf
    \State $H_t[C'[j]] \gets H_t[C'[j]] + w$
  \EndFor
  \Return{$H_t$}
\EndFunction
\end{algorithmic}
\end{algorithm}

In each iteration $l_i$ of the algorithm, we toggle the \textit{Pick-Less (PL)} mode to mitigate community swaps. The PL mode is enabled every $\rho$ iterations (lines \ref{alg:vlouvainlm--iterations-begin}-\ref{alg:vlouvainlm--init-deltaq}), ensuring that a vertex can only switch to a new community with a smaller ID during specific iterations. At the start of each iteration, the total change in modularity, $\Delta Q$, is initialized to zero. Within each iteration, all unprocessed vertices in $G'$ are handled in parallel (lines \ref{alg:vlouvainlm--loop-vertices-begin}-\ref{alg:vlouvainlm--loop-vertices-end}). For a given vertex $i$, we first mark it as processed to prune unnecessary computations (line \ref{alg:vlouvainlm--prune}). Depending on the degree of vertex $i$, either an individual thread or a thread-block is assigned to process it. A hashtable $H$ is allocated from buffers $buf_k$ and $buf_v$ to track the labels of neighboring communities (lines \ref{alg:vlouvainlm--loop-vertices-begin}-\ref{alg:vlouvainlm--scan}). The hashtable is initialized in parallel, and each neighbor $(j, w)$ of vertex $i$ accumulates its community label $C[j]$ and edge weight $w$ to the hashtable using the \texttt{hashtableAccumulate()} function. Once all neighbors of $i$ are processed, we identified the best community $c^*$ for $i$ using the aggregated weights in $H$ (lines \ref{alg:vlouvainlm--best-community-begin}-\ref{alg:vlouvainlm--best-community-end}). The delta-modularity $\delta Q^*$ resulting from moving $i$ to $c^*$ is calculated, and $i$ adopts the new label $c^*$ only if it improves modularity and satisfies the PL mode constraints. This involves updating the total edge weights $\Sigma'$ of the old and new communities (line \ref{alg:vlouvainlm--perform-move-begin}) and updating $C'[i]$ in place. The neighbors of $i$ are then marked as unprocessed to allow subsequent refinement (line \ref{alg:vlouvainlm--remark}). At the end of each iteration, the algorithm checks if the modularity improvement $\Delta Q$ is below the tolerance $\tau$. If so, the iterations terminate, as the communities have locally converged (line \ref{alg:vlouvainlm--locally-converged}). Finally, we return the number of iterations performed in the local-moving phase, $l_i$, in line \ref{alg:vlouvainlm--return}.

The \texttt{scanCommunities} function, invoked during the neighbor traversal phase, aggregates weights of neighboring community labels for a vertex $i$. It ensures that the contributions of self-loops are appropriately excluded when the \texttt{self} flag is disabled (lines \ref{alg:vlouvainlm--loop-vertices-begin}-\ref{alg:vlouvainlm--scan}). This function enables efficient computation of the best community for $i$ while maintaining high parallelism.

\subsubsection{Aggregation phase of $\nu$-Louvain}

The psuedocode for the aggregation phase of $\nu$-Louvain is given in Algorithm \ref{alg:vlouvainag}. It constructs a super-vertex graph $G''$ from the input graph $G'$ and the community memberships $C'$ identified during the local-moving phase.

\begin{algorithm}[hbtp]
\caption{Aggregation phase of $\nu$-Louvain.}
\label{alg:vlouvainag}
\begin{algorithmic}[1]
\Require{$G'$: Input/super-vertex graph}
\Require{$C'$: Community membership of each vertex}
\Ensure{$G'_{C'}$: Community vertices (CSR)}
\Ensure{$G''$: Super-vertex graph (weighted CSR)}
\Ensure{$*.offsets$: Offsets array of a CSR graph}
\Ensure{$H$: Per-vertex hashtable ($H_k$: keys, $H_v$: values)}
\Ensure{$p_1$: Capacity of $H$, and also a prime}
\Ensure{$p_2$: Secondary prime, such that $p_2 > p_1$}

\Statex

\Function{$\nu$louvainAggregate}{$G', C'$}
  \State $\rhd$ Obtain vertices belonging to each community
  \State $G'_{C'}.offsets \gets countCommunityVertices(G', C')$ \label{alg:vlouvainag--coff-begin}
  \State $G'_{C'}.offsets \gets exclusiveScan(G'_{C'}.offsets)$ \label{alg:vlouvainag--coff-end}
  \ForAll{$i \in V'$ \textbf{in parallel}} \label{alg:vlouvainag--comv-begin}
    \State Add edge $(C'[i], i)$ to CSR $G'_{C'}$ atomically
  \EndFor \label{alg:vlouvainag--comv-end}
  \State $\rhd$ Obtain super-vertex graph
  \State $G''.offsets \gets communityTotalDegree(G', C')$ \label{alg:vlouvainag--yoff-begin}
  \State $G''.offsets \gets exclusiveScan(G''.offsets)$ \label{alg:vlouvainag--yoff-end}
  \State $|\Gamma| \gets$ Number of communities in $C'$
  \ForAll{$c \in [0, |\Gamma|)$ \textbf{in parallel}} \label{alg:vlouvainag--y-begin}
    \If{degree of $c$ in $G'_{C'} = 0$} \textbf{continue}
    \EndIf
    \State $\rhd$ If degree of $c < \text{\small{SWITCH\_DEGREE\_AGGREGATE}}$,
    \State $\rhd$ process using a thread, else use a thread-block
    \State $p_1 \gets nextPow2(G'_{C'}.degree(c)) - 1$
    \State $p_2 \gets nextPow2(p_1) - 1$
    \State $\theta_H \gets 2 * G'_{C'}.offset(c)$
    \State $H_k \gets buf_k[\theta_H : \theta_H + p_1]$ \Comment{$H$ is \textbf{shared}, if using}
    \State $H_v \gets buf_v[\theta_H : \theta_H + p_1]$ \Comment{a thread-block}
    \State $hashtableClear(H)$ \textbf{in parallel}
    \ForAll{$i \in G'_{C'}.neighbors(c)$} \textbf{in parallel}
      \ForAll{$(j, w) \in G'.neighbors(i)$}
        \State $hashtableAccumulate(H, p_1, p_2, C[j], w)$
      \EndFor
    \EndFor
    \ForAll{$(d, w) \in H$} \textbf{in parallel}
      \State Add edge $(c, d, w)$ to CSR $G''$ atomically
    \EndFor
  \EndFor \label{alg:vlouvainag--y-end}
  \Return $G''$ \label{alg:vlouvainag--return}
\EndFunction
\end{algorithmic}
\end{algorithm}

The first step of the algorithm is to group vertices into their respective communities. Lines \ref{alg:vlouvainag--coff-begin}-\ref{alg:vlouvainag--coff-end} compute the offsets for the community vertices in a Compressed Sparse Row (CSR) representation of $G'_{C'}$. Specifically, the function \texttt{countCommunityVertices()} determines the count of vertices for each community, and an exclusive scan is performed to convert these counts into offsets. In lines \ref{alg:vlouvainag--comv-begin}-\ref{alg:vlouvainag--comv-end}, each vertex $i$ is added to its corresponding community $C'[i]$ atomically, populating $G'_{C'}$. Next, we construct the super-vertex graph $G''$, where communities act as vertices and inter-community edges are aggregated. Lines \ref{alg:vlouvainag--yoff-begin}-\ref{alg:vlouvainag--yoff-end} calculate the total degree for each community and determine offsets for the CSR representation of $G''$. The total number of communities, $|\Gamma|$, is computed, and the algorithm processes each community $c$ in parallel (lines \ref{alg:vlouvainag--y-begin}-\ref{alg:vlouvainag--y-end}). For each community $c$, we determine whether to process it using a single thread or a thread-block, based on its degree compared to \texttt{\small{SWITCH\_DEGREE\_AGGREGATE}}. Memory for the hashtable $H$ is allocated from shared buffers $buf_k$ and $buf_v$, with $H_k$ and $H_v$ representing the key and value arrays, respectively. The hashtable is cleared in parallel. Next, for each neighbor $i$ of community $c$ in $G'_{C'}$, we iterate over the neighbors $(j, w)$ of $i$ in the original graph $G'$, and accumulate $(C[j], w)$ into the hashtable $H$ (lines \ref{alg:vlouvainag--y-begin}-\ref{alg:vlouvainag--y-end}). After processing neighbors of all vertices in community $c$, a parallel iteration over $H$ constructs the edges $(c, d, w)$, representing the aggregated inter-community weights, which are added atomically to the CSR $G''$. Finally, we return the super-vertex graph $G''$ in line \ref{alg:vlouvainag--return}.

\begin{algorithm}[hbtp]
\caption{Accumulating associated weights of labels \cite{sahu2024nulpa}.}
\label{alg:vhashtable}
\begin{algorithmic}[1]
\Require{$H$: Hashtable with $p_1$ slots ($H_k$: keys, $H_v$: values)}
\Require{$p_1$: Capacity of $H$, and also a prime}
\Require{$p_2$: Secondary prime, such that $p_2 > p_1$}
\Require{$k, v$: Key, associated value to accumulate}
\Ensure{$s$: Slot index}

\Statex

\Function{hashtableAccumulate}{$H, p_1, p_2, k, v$}
  \State $i \gets k$ \textbf{;} $\delta i \gets 1$ \label{alg:vhashtable--init}
  \ForAll{$t \in [0\ \dots\ \text{\small{MAX\_RETRIES}})$} \label{alg:vhashtable--loop-begin}
    \State $s \gets i \bmod p_1$ \Comment{$1^{st}$ hash function} \label{alg:vhashtable--slot}
    \If{\textbf{is not shared}}
      \If{$H_k[s] = k$ \textbf{or} $H_k[s] = \phi$} \label{alg:vhashtable--shared-begin}
        \If{$H_k[s] = \phi$} $H_k[s] \gets k$
        \EndIf
        \State $H_v[s] \gets v$
        \Return{$done$}
      \EndIf \label{alg:vhashtable--shared-end}
    \Else
      \If{$H_k[s] = k$ \textbf{or} $H_k[s] = \phi$} \label{alg:vhashtable--unshared-begin}
        \State $k_{old} \gets atomicCAS(H_k[s], \phi, k)$
        \If{$k_{old} = \phi$ \textbf{or} $k_{old} = k$}
          \State $atomicAdd(H_v[s], v)$
          \Return{$done$}
        \EndIf
      \EndIf \label{alg:vhashtable--unshared-end}
    \EndIf
    \State $i \gets i + \delta i$ \label{alg:vhashtable--update-begin}
    \State $\delta i \gets 2 * \delta i + (k \bmod p_2)$ \Comment{$2^{nd}$ hash function}\label{alg:vhashtable--update-end}
  \EndFor \label{alg:vhashtable--loop-end}
  \Return{$failed$}
\EndFunction
\end{algorithmic}
\end{algorithm}

\subsubsection{Hashtable for our GPU-based Louvain}

Algorithm \ref{alg:vhashtable} gives the psuedocode for accumulating the associated weights of neighboring communities for a vertex in its hashtable while resolving collisions efficiently using a hybrid quadratic-double probing strategy. Here, the goal is to find a suitable slot in the hashtable $H$ (with $p_1$ slots) to either insert a given key $k$ with its value $v$, or add $v$ to an existing entry. The algorithm starts by calculating an initial slot index $s$ using the primary hash function: $s = i \bmod p_1$ (line \ref{alg:vhashtable--slot}), where $i$ is initialized to $k$ and adjusted incrementally with a step size $\delta i$. If the slot $s$ is empty or contains the key $k$, we insert or update the value at $H_v[s]$. The behavior depends on whether the hashtable is accessed by multiple threads (shared) or a single thread (non-shared). In the non-shared case (lines \ref{alg:vhashtable--unshared-begin}-\ref{alg:vhashtable--unshared-end}), $H_k$ and $H_v$ are modified directly. For the shared case (lines \ref{alg:vhashtable--shared-begin}-\ref{alg:vhashtable--shared-end}), atomic operations ensure thread safety. Specifically, an \texttt{atomicCAS()} operation atomically sets $H_k[s]$ to $k$ if the slot is empty. If successful, or if the slot already contains $k$, an \texttt{atomicAdd()} operation updates $H_v[s]$ by adding $v$. When the slot is occupied by a different key, we use hybrid quadratic-double probing: $\delta i$ is doubled and adjusted with $k \bmod p_2$ (line \ref{alg:vhashtable--update-end}), where $p_2$ is a secondary prime larger than $p_1$. This process iterates up to \texttt{\small{MAX\_RETRIES}} (lines \ref{alg:vhashtable--loop-begin}-\ref{alg:vhashtable--loop-end}). If no suitable slot is found within the allowed attempts, a \textit{failed} status is returned. However, this is avoided by ensuring the hashtable is appropriately sized\ignore{to accommodate all entries}.

\end{document}